\documentclass[aps, prd, twocolumn, preprintnumbers, groupedaddress, nofootinbib, amssymb, notitlepage, eqsecnum]{revtex4-2}
\usepackage{bm}
\usepackage[dvipsnames]{xcolor}
\usepackage{amsmath,amsthm,amssymb}
\usepackage{amsfonts}
\usepackage{hyperref}
\usepackage{graphicx}

\allowdisplaybreaks[1]

\newcommand{\rd}{{\rm d}}
\newcommand{\be}{\begin{equation}}
\newcommand{\ee}{\end{equation}}
\newcommand{\ba}{\begin{eqnarray}}
\newcommand{\ea}{\end{eqnarray}}
\newcommand{\Mpl}{M_{\rm Pl}}

\begin{document}

\preprint{YITP-25-178, WUCG-25-13}

\title{Exotic compact objects in Einstein-scalar-Maxwell theories}

\author{Antonio De Felice$^{a}$}
\email{antonio.defelice@yukawa.kyoto-u.ac.jp}  
\author{Shinji Tsujikawa$^{b}$}
\email{tsujikawa@waseda.jp} 

\affiliation{$^{a}$Center for Gravitational Physics and Quantum Information, Yukawa Institute for Theoretical Physics, Kyoto University, 606-8502, Kyoto, Japan\\
$^{b}$Department of Physics, Waseda University, 
3-4-1 Okubo, Shinjuku, Tokyo 169-8555, Japan}
 
\date{\today}

\begin{abstract}

In k-essence theories within general relativity, where the matter Lagrangian depends on a real scalar field $\phi$ and its kinetic term $X$, static and spherically symmetric compact objects with a positive-definite energy density cannot exist without introducing ghosts. We show that this no-go theorem can be evaded when the k-essence Lagrangian is extended to include a dependence on the field strength $F$ of a $U(1)$ gauge field, taking the general form ${\cal L}(\phi, X, F)$. In Einstein-scalar-Maxwell theories with a scalar-vector coupling $\mu(\phi)F$, we demonstrate the existence of asymptotically flat, charged compact stars whose energy density and pressure vanish at the center. With an appropriate choice of the coupling function $\mu(\phi)$, we construct both electric and magnetic compact objects and derive their metric functions and scalar- and vector-field profiles analytically. 
We compute their masses and radii, showing that the compactness lies in the range 
${\cal O}(0.01)<{\cal C}<{\cal O}(0.1)$. 
A linear perturbation analysis reveals that electric compact objects are free of strong coupling, ghost, and Laplacian instabilities at all radii for $\mu(\phi)>0$, while magnetic compact objects suffer from 
strong coupling near the center.

\end{abstract}


\maketitle

\section{Introduction}
\label{Intro}

The advent of gravitational-wave astronomy 
\cite{LIGOScientific:2016aoc,LIGOScientific:2017vwq} and very long baseline interferometry \cite{Doeleman:2008qh,EventHorizonTelescope:2019ggy} has opened new windows for probing physics in strong-gravity regimes. 
Black holes (BHs) are extremely compact objects with an event horizon, predicted as vacuum solutions \cite{Schwarzschild:1916uq,Kerr:1963ud} to the Einstein equations in general relativity (GR) \cite{Einstein:1916vd}. 
Neutron stars (NSs) are also compact objects without an event horizon, whose geometries are determined by including a baryonic matter source in Einstein equations, typically modeled as a perfect fluid.
In addition to the presence or absence of the horizon, NSs have regular centers, whereas BHs typically feature singularities at their centers. The properties of BHs 
and NSs can be constrained by observations of gravitational waves emitted from their binary systems.

Cosmological observations indicate that about 95\,\% of the energy density in the present Universe consists of dark energy and dark matter \cite{Planck:2018vyg,Scolnic:2021amr,DES:2024jxu,DESI:2024mwx,DESI:2025zgx}. 
While dark energy drives cosmic acceleration on large scales, dark matter can gravitationally cluster to form the large-scale structure 
of the Universe \cite{Peebles:1982ff,Peebles:1984ge}.
Possible candidates for dark matter include 
(pseudo)-scalar fields such as axions \cite{Peccei:1977hh,Weinberg:1977ma,Wilczek:1977pj,Kim:1979if,Shifman:1979if,Marsh:2015xka}, as well as vector fields such as dark photons \cite{Holdom:1985ag,Fayet:1990wx,Arkani-Hamed:2008hhe,Pospelov:2008jd,Fabbrichesi:2020wbt}. 
The usual horizonless compact 
objects are composed of baryonic matter, 
but there are also possibilities that 
scalar or vector fields give rise to 
self-gravitating exotic compact 
objects (ECOs) \cite{Schunck:2003kk,Liebling:2012fv,Cardoso:2019rvt}.

One of the well-known examples of such ECOs is a boson star composed of a complex scalar field with a potential \cite{Kaup:1968zz,Ruffini:1969qy}. 
For a real canonical scalar field $\phi$ 
with the kinetic term 
$X=-(1/2)\partial_{\mu}
\phi \partial^{\mu}\phi$ and potential $V(\phi)$, which is given by the Lagrangian 
${\cal L} = X - V(\phi)$, it is known that static and spherically symmetric (SSS) 
configurations with a positive-definite energy density do not exist within the 
framework of GR
\cite{Derrick:1964ww}. 
This property also holds for noncanonical scalar fields described by a k-essence Lagrangian ${\cal L}(\phi, X)$, as long as the ghost-free condition $\partial{\cal L}/\partial X > 0$ is 
satisfied \cite{Diez-Tejedor:2013sza}.
This is why the real scalar field is extended to a complex scalar field to realize the SSS configuration 
of boson stars.

In the case of vector fields, 
Wheeler {\it et al.}~\cite{Wheeler:1955zz,Power:1957zz} originally attempted to construct 
particle-like solutions, known as 
``geons,'' by studying electromagnetism within the framework of GR. 
However, these geons were found to be unstable. By considering a complex Abelian vector field with a mass term, it is possible to form gravitational solitons, known as Proca stars \cite{Brito:2015pxa,Herdeiro:2016tmi}, within the framework of GR. Such a massive vector field can be associated with dark photons, i.e., 
one of the candidates for dark matter. 
It is also possible to extend Maxwell’s theory with a real vector field to nonlinear electrodynamics, described by a Lagrangian of the form ${\cal L}(F)$ \cite{Heisenberg:1936nmg,Born:1934gh}, where $F$ denotes the gauge-field field strength.
By appropriately designing the functional form of ${\cal L}(F)$, one can construct SSS solutions that are regular at the center ($r=0$), including nonsingular 
BHs \cite{Ayon-Beato:1998hmi,Ayon-Beato:1999kuh,Ayon-Beato:2000mjt,Bronnikov:2000vy,
Dymnikova:2004zc}.
However, it has been shown that such solutions are unstable due to a Laplacian instability along the angular direction near $r=0$ \cite{DeFelice:2024seu}.

Given that real scalar and vector fields face difficulties in realizing SSS ECOs 
without instabilities, an important question is what occurs in theories containing both 
fields---particularly interactions 
between the scalar and vector fields.
For example, one can consider 
Einstein-Scalar-Maxwell (ESM) theories 
described by the Lagrangian 
$\Mpl^2 R/2+X+\mu(\phi)F$, 
where $\Mpl$ is the reduced Planck mass, $R$ is the Ricci scalar, and $\mu(\phi)$ is the scalar-vector coupling that 
depends on the scalar field 
$\phi$ \cite{Gaillard:1981rj,Duff:1995sm,Andrianopoli:1996cm}. 
For a dilatonic coupling of the form $\mu(\phi) = \mu_0 e^{-\lambda \phi}$, there exists an exact SSS BH solution endowed with scalar and electric (or magnetic) charges \cite{Gibbons:1987ps,Garfinkle:1990qj}, 
featuring curvature singularities at $r=0$. By considering alternative forms of $\mu(\phi)$, it may be possible to realize star-like configurations that are free of singularities at their centers (see Ref.~\cite{Cardoso:2021ehg} for a charged boson star composed of a complex scalar field coupled to a real vector field).

Indeed, in theories described by the 
Lagrangian $\Mpl^2 R/2+{\cal L}(\phi,X,F)$, 
the possibility of realizing compact objects 
with regular centers was 
studied in Ref.~\cite{DeFelice:2024ops}.
By examining several classes of theories encompassed by the matter Lagrangian 
${\cal L}(\phi,X,F)$, no nonsingular BHs free of ghost or Laplacian instabilities have been found. However, there exist particular theories that allow the existence of horizonless compact objects without such instabilities. 
This includes ESM theories with the matter Lagrangian ${\cal L}=
X + \mu(\phi) F$ mentioned above. 
Under the condition $\mu(\phi) > 0$, ghosts are absent, and the propagation speeds are luminal in both the radial and angular directions (see Refs.~\cite{Heisenberg:2018mgr,Gannouji:2021oqz,Kase:2023kvq} for the 
analysis of perturbations in more general 
scalar-vector-tensor theories).

In Ref.~\cite{Herdeiro:2019oqp}, it was shown 
that regular and asymptotically flat compact star 
solutions do not exist in ESM theories when the 
coupling $\mu(\phi)$ remains bounded at $r = 0$.
This no-go theorem can be circumvented 
if the coupling diverges at $r = 0$,
while all physical quantities remain finite \cite{Herdeiro:2019iwl}.
Indeed, this latter situation is realized in the electrically charged compact
object proposed in Ref.~\cite{DeFelice:2024ops}.

Despite the possible existence of regular compact objects 
in ESM theories~\cite{Herdeiro:2019iwl,DeFelice:2024ops}, 
the profiles of the energy density and pressure 
have not yet been elucidated. 
In particular, it remains unclear how the no-go theorem for k-essence can be circumvented by incorporating the $F$ dependence coupled to the scalar field. 
In this paper, we address this issue by computing the energy density, as well as the radial and transverse pressures, as functions of the radial coordinate $r$. To this end, we consider the case in which the ratio $N(r)$ between the two metric functions is known as a function of $r$. We then derive analytic solutions for the metric functions, as well as for the scalar and vector field profiles. 

In such a scenario, the same profiles of energy density and pressure are realized for both electric and magnetic objects, under appropriate choices of $\mu(\phi)$. In the magnetic case, the form of $\mu(\phi)$ is inversely related to that in the electric case, so that its functional dependence on $\phi$ differs between the two cases.
We show that the energy density, which is positive for all $r>0$, attains its maximum at an intermediate radius, while it vanishes at both $r = 0$ and spatial infinity. We also plot the mass-radius relation by defining the 
radius $r_s$ at which the mass function reaches 99\,\% 
of the total ADM mass $M$. For a given length scale $r_0$, where each compact object is characterized by its own value of $r_0$ appearing in the function $N(r)$, we find that $M$ increases with $r_s$ and asymptotically approaches the maximum value $M_{\rm max}$. 
The compactness of charged objects can exceed 
the order of 0.01. 
We also find a universal relation between 
the ADM mass and the stellar radius, since $M/r_s$ is fixed by the parameters of the Lagrangian once the criterion for determining the radius $r_s$ is specified, 
as mentioned above.

While the ECOs realized in our models are free of ghost and Laplacian instabilities in both the electric and magnetic cases, it remains unclear whether such stars suffer from the strong coupling problem. Since this issue was not addressed in Ref.~\cite{DeFelice:2024ops}, we investigate it by considering perturbations around the SSS background. For the electric compact object, the coupling $\mu(\phi)$ diverges as $r \to 0$, and hence the vector-field perturbation is weakly coupled. In contrast, for the magnetic object, $\mu(\phi)$ approaches 0 as $r \to 0$, so that the vector-field perturbation becomes strongly coupled. 
Since nonlinear perturbations 
are out of control in the latter case, 
the background profile for the magnetic 
object loses its validity near the center. 
Therefore, the electric compact object
is only the case in which the strong 
coupling problem is absent at any 
distance $r$. 
Even though $\mu(\phi)$ diverges at $r=0$, 
the interaction term $\mu(\phi) F$ is proportional 
to $r^{2}$, and hence no divergence appears 
in physical quantities.
Thus, linearly stable electric ECOs free 
from the strong coupling problem exist in ESM theories.

This paper is organized as follows. 
In Sec.~\ref{backsec}, we derive the field equations on the SSS background for theories described by the Lagrangian $\Mpl^2 R/2 + {\cal L}(\phi,X,F)$, and show how the no-go theorem for k-essence can be circumvented. 
In Sec.~\ref{electricsec}, we study the configurations of electric compact objects realized in ESM theories, with particular attention to the profiles of the energy density, radial and angular pressures, and the mass-radius relation. 
In Sec.~\ref{magneticsec}, we show that the same density profile and mass-radius relation as those obtained for the electric object can be realized for the magnetic object by choosing the coupling $\mu(\phi)$ to be inversely proportional to that in the electric case. In Sec.~\ref{obsersec}, we address the issue of strong coupling and show that only the electric compact object is free from this problem. Sec.~\ref{consec} is devoted to the conclusions.

\section{Field equations on the 
SSS background}
\label{backsec}

We aim to construct nonsingular horizonless SSS 
objects with regular centers and study their properties.
For this purpose, we introduce a real scalar field $\phi$ with the kinetic term $X 
=-(1/2)\partial_{\mu}\phi \partial^{\mu}\phi$, and a $U(1)$ gauge field $A_{\mu}$ characterized by the 
field strength $F = -(1/4)F_{\mu\nu}F^{\mu\nu}$, where $F_{\mu\nu} = \partial_{\mu}A_{\nu} - \partial_{\nu}A_{\mu}$. 
We assume that these scalar and vector fields are present in the dark sector of the Universe.
The gravitational sector is described by GR, whose dynamics is governed by the 
Einstein-Hilbert Lagrangian $\Mpl^{2}R/2$. 
The total action is given by 
\be
{\cal S}=\int \rd^4 x \sqrt{-g} 
\left[ \frac{\Mpl^2}{2}R
+{\cal L}(\phi, X, F) \right]\,,
\label{action}
\ee
where $g$ denotes the determinant of the metric tensor $g_{\mu\nu}$, and ${\cal L}$ represents a function of 
$\phi$, $X$, and $F$.
The covariant equations of motion obtained by varying Eq.~(\ref{action}) with respect to $g^{\mu\nu}$ take the form 
\be
\Mpl^2 {G^{\mu}}_{\nu}
={T^{\mu}}_{\nu}\,,
\label{graeq}
\ee
where ${G^{\mu}}_{\nu}$ is the Einstein tensor and ${T^{\mu}}_{\nu}$ denotes the energy-momentum tensor of matter, expressed as
\be
{T^{\mu}}_{\nu}=
{\cal L}\,{\delta^{\mu}}_{\nu}
+{\cal L}_{,X} \partial^{\mu}\phi 
\partial_{\nu}\phi
+{\cal L}_{,F} {F^{\mu}}_{\lambda} 
{F_{\nu}}^{\lambda}\,,
\ee
with  
${\cal L}_{,X} \equiv \partial{\cal L}/
\partial X$ and 
${\cal L}_{,F} \equiv \partial{\cal L}/
\partial F$.
 
We take a SSS background 
described by the line element
\be
\rd s^{2} =-f(r) \rd t^{2}
+h^{-1}(r)\rd r^{2} + r^{2} (\rd \theta^{2}
+\sin^{2}\theta\, \rd \varphi^{2})\,,
\label{metric}
\ee
where $f$ and $h$ depend on the radial distance $r$. 
For horizonless compact objects, both $f(r)$ and $h(r)$ remain positive for all values of $r$. 
On this background, we consider the scalar field $\phi$ to depend only on the radial coordinate $r$.
The theory (\ref{action}) is invariant under the gauge transformation 
$A_{\mu} \to A_{\mu} + \partial_{\mu}\chi$, which ensures that the vector field has no longitudinal mode. 
In this case, we can choose the 
following configuration for $A_{\mu}$:
\be
A_{\mu} \rd x^\mu=A_0(r)\rd t 
-q_M \cos \theta\,\rd\varphi\,,
\label{Amu}
\ee
where $A_0(r)$ is a function of the radial coordinate $r$, and $q_M$ is a constant associated with the magnetic 
charge. For this configuration, 
the scalar quantities $X$ and $F$ 
can be written as
\be
X=-\frac{1}{2}h \phi'^2\,,\qquad 
F=\frac{h A_0'^2}{2f} 
-\frac{q_M^2}{2r^4}\,,
\label{XF}
\ee
where a prime denotes a derivative with respect to the radial coordinate $r$.

From the (00), (11), and 
(22) [=(33)] components of 
the gravitational 
equation (\ref{graeq}), we obtain
\ba
& &
\frac{\Mpl^2 (rh'+h-1)}{r^2}
=-\rho\,,\label{back1} \\
& &
\frac{\Mpl^2 [rhf'+f(h-1)]}{r^2 f}=P_r\,,\label{graeq2} \\
& &
\frac{\Mpl^2 [2rfh f'' 
-rh f'^2+ ( rh' + 2h) ff' 
+ 2f^2h']}{4rf^2} 
\nonumber \\
& &
=P_{t}\,, 
\label{graeq3}
\ea
where 
\ba
\rho &\equiv& -{T^{0}}_0
=\frac{h A_0'^2}{f} 
{\cal L}_{,F}
-{\cal L}\,,\label{rho} \\
P_r &\equiv& {T^{1}}_1=
{\cal L}+h\phi'^2 {\cal L}_{,X}
-\frac{h A_0'^2}{f} 
{\cal L}_{,F}\,,\label{Pr} \\
P_{t} &\equiv& 
{T^{2}}_2=
{\cal L}+\frac{q_M^2 {\cal L}_{,F}}{r^4}\,.\label{Pt}
\ea
Here, $\rho$ denotes the energy density sourced by $\phi$ and $A_{\mu}$, while $P_r$ and $P_t$ correspond to the radial and transverse pressures, respectively.
In contrast to a perfect fluid, $P_r$ and $P_t$ are generally not equal.

We define the mass function ${\cal M}(r)$ as
\be
{\cal M}(r) \equiv 
4\pi \Mpl^2 r \left[ 1-h(r) \right]\,.
\label{mfunction}
\ee
From Eq.~(\ref{back1}), we then 
obtain the differential equation
\be
{\cal M}'(r)=4\pi \rho(r)\,r^2\,,
\label{Mdiff}
\ee
whose solution can be expressed in the integrated form 
${\cal M}(r)=\int_0^r 4\pi \rho(\tilde{r})\,
\tilde{r}^2 {\rm d}\tilde{r}$. 
The Arnowitt-Deser-Misner (ADM) mass of the 
SSS object is 
defined by the limit
\be
M=\lim_{r \to \infty}
4\pi \Mpl^2 r \left[ 1-h(r) \right]\,,
\label{ADM}
\ee
where the compactness of the object 
implies that $M$ is constant.

Using the continuity equation, $\nabla_{\mu} {T^{\mu}}_{\nu} = 0$, 
we obtain the following relation between the energy density and the pressures:
\be
P_r'+\frac{f'}{2f} 
\left( \rho+P_r \right)
+\frac{2}{r}
\left( P_r-P_{t} \right)=0\,,
\label{continuity}
\ee
which can also be derived by combining 
Eqs.~(\ref{back1})-(\ref{graeq3}).
The difference between $P_r$ and $P_t$ provides a modification to the 
continuity equation of a 
perfect fluid, $P_r'+f'(\rho+P_r)/(2f)=0$.
Using Eqs.~(\ref{rho})-(\ref{Pt}), 
Eq.~(\ref{continuity}) can be 
written as
\be
P_r'+\left( \frac{f'}{2f}
+\frac{2}{r} \right) h \phi'^2 
{\cal L}_{,X}
-\frac{2}{r} \left( 
\frac{h A_0'^2}{f}
+\frac{q_M^2}{r^4} \right) 
{\cal L}_{,F}=0\,.
\label{continuity2}
\ee

In k-essence theories, which are characterized by the Lagrangian ${\cal L}(\phi,X)$ \cite{Armendariz-Picon:1999hyi,Chiba:1999ka,Armendariz-Picon:2000nqq}, the third term on the left-hand side of Eq.~(\ref{continuity2}) vanishes due to the property ${\cal L}_{,F} = 0$. 
The absence of ghosts requires the condition ${\cal L}_{,X}>0$. 
The gravitational force is attractive when $f'/f > 0$, so that $P_r'(r) < 0$ for ${\cal L}_{,F} = 0$.
By applying the boundary condition $P_r(r \to \infty) = 0$, we obtain $P_r(r) > 0$ under the condition $P_r'(r) < 0$.
At $r = 0$, we impose the condition $\phi'(0) = 0$ to avoid the appearance of a cusp-like structure that would violate the SSS configuration.
It then follows that $\rho(r=0)=-P_r(r=0)<0$. This means that 
the energy density of the 
scalar field is negative. 
Thus, in k-essence theories satisfying the no-ghost condition, there exist no SSS objects with a positive energy density at the center \cite{Diez-Tejedor:2013sza}. This no-go theorem can be circumvented by including the $F$ dependence in the k-essence Lagrangian.

We introduce the function 
\be
N(r)=\frac{f(r)}{h(r)}\,,
\ee
which is positive to ensure 
the positivity of $-g$. 
Combining Eq.~(\ref{graeq2}) with Eq.~(\ref{back1}), we find 
\be
\frac{N'}{N}=
\frac{r \phi'^2{\cal L}_{,X}}{\Mpl^2}\,.
\label{back2}
\ee
The $X$ dependence in ${\cal L}$ implies 
$N' \neq 0$, resulting in a difference between $f(r)$ and $h(r)$. If ${\cal L}$ does not depend on $X$, we have $N' = 0$, so that $N(r) = 1$ by imposing the boundary condition $f(r) = h(r) \to 1$ as $r \to \infty$.

Varying the action (\ref{action}) with respect to $A_0$ and $\phi$, 
we obtain 
\ba
& &
A_0'=\frac{q_E \sqrt{N}}{r^2 {\cal L}_{,F}}\,,
\label{back3}\\
& &
\left( \sqrt{N} h r^2 \phi'{\cal L}_{,X} \right)'
+\sqrt{N}r^2 {\cal L}_{,\phi}=0\,,
\label{back4}
\ea
where $q_E$ is an integration constant corresponding to the electric charge, and 
${\cal L}_{,\phi} \equiv \partial {\cal L}/\partial \phi$.
For a purely magnetically charged object ($q_E = 0$), the electric field $A_0'$ vanishes, as expected.  
Equations (\ref{back1}), (\ref{back2}), (\ref{back3}), and (\ref{back4}) govern the background dynamics of our system. We note that Eq.~(\ref{graeq3}) can be obtained by differentiating Eq.~(\ref{back2}) and using the other equations of motion.

The energy density can be expressed as 
$\rho=A_0' q_E/(r^2 
\sqrt{N})-{\cal L}$. 
By taking the derivative of Eq.~(\ref{back1}) with respect to $r$, using the relation 
${\cal L}'=
{\cal L}_{,\phi}\phi'
+{\cal L}_{,X}X'
+{\cal L}_{,F}F'$, 
and eliminating ${\cal L}_{,F}$, 
${\cal L}_{,X}$, ${\cal L}_{,\phi}$ terms 
by using Eqs.~(\ref{back1}), (\ref{back2}), 
and (\ref{back4}), we obtain
\ba
& &
4q_E r^2 A_0'^2-\Mpl^2 r^2 N^{-3/2} 
[2N^2 (r^2 h''-2h+2) 
\nonumber \\
& &
+r N (2r h N''+3r h' N'+2 h N')
-r^2 h N'^2]A_0'
\nonumber \\
& &
+4 N q_E q_M^2/r^2=0\,.
\label{A0con}
\ea

For SSS objects without curvature singularities at their center, 
the metric functions near $r=0$ 
must take the forms
\ba
h(r) &=& 1+\sum_{n=2}^{\infty}h_n r^n\,,\label{hr}\\
N(r) &=& N_0+\sum_{n=2}^{\infty}N_n r^n\,,\label{Nr}
\ea
where $h_n$, $N_0$, and $N_n$ are constants. 
Substituting Eqs.~(\ref{hr}) and (\ref{Nr}) into Eq.~(\ref{A0con}) and expanding around $r = 0$, the leading-order term of $A_0'$ is given by  
$A_0'=\sqrt{N_0}
\sqrt{-(q_E q_M)^2}
/(q_E r^2)$. 
Hence, a consistent solution exists only if $q_E q_M = 0$ \cite{DeFelice:2024ops}. 
Therefore, we focus on either the purely electric object ($q_E \neq 0$, 
$q_M = 0$) or the purely magnetic object 
($q_M \neq 0$, $q_E = 0$).

In this paper, we study the properties of ECOs in ESM theories, where the matter sector is described 
by the Lagrangian
\be
{\cal L}=X+\mu (\phi)F\,,
\label{model}
\ee
with $\mu$ being a function of $\phi$. In this case, Eqs.~(\ref{rho}), (\ref{Pr}), 
and (\ref{Pt}) reduce, respectively, to
\ba
\rho &=& \frac{1}{2}h \phi'^2
+\frac{\mu A_0'^2}{2N}
+\frac{\mu q_M^2}{2r^4}\,,
\label{rhoD} \\
P_r &=& \frac{1}{2}h \phi'^2
-\frac{\mu A_0'^2}{2N}
-\frac{\mu q_M^2}{2r^4}\,,\\
P_{t} &=& -P_r\,.
\label{PtD}
\ea
It should be noted that using the relation $P_t + P_r = 0$ in Eqs.~\eqref{graeq2} and \eqref{graeq3} yields a constraint equation for the metric functions.
The solutions that we study in the following 
sections automatically satisfy this constraint.

Ref.~\cite{DeFelice:2024ops} studied linear perturbations around the background (\ref{action}) and derived the stability conditions against odd- and even-parity perturbations for theories described by the action (\ref{action}).
For both electric and magnetic objects, the absence of ghosts requires that 
\be
{\cal L}_{,X}>0\,,\qquad 
{\cal L}_{,F}>0\,.
\ee
Since ${\cal L}_{,X} = 1$ in ESM theories, the first condition is trivially satisfied. The second 
condition reduces to 
\be
{\cal L}_{,F}=\mu (\phi) >0\,.
\label{mucon}
\ee
The Lagrangian (\ref{model}) contains only linear dependence on $X$ and $F$, so that ${\cal L}_{,XX} = 0$, ${\cal L}_{,FF} = 0$, and ${\cal L}_{,XF} = 0$. In this case, the propagation speeds of odd- and even-parity perturbations along both radial and angular directions are equal to the speed of light for both electric and magnetic objects \cite{DeFelice:2024ops}. This implies the absence of Laplacian instabilities in ESM theories.

Under the condition (\ref{mucon}), the energy density given in Eq.~(\ref{rho}) is positive. We also note that the third term on the left-hand side of Eq.~(\ref{continuity2}) is negative, whereas the second term is positive for $f'/f>0$. 
This allows for a region in which $P_r'(r)>0$, so that $P_r(r)$ can become negative under the boundary condition $P_r (r \to \infty)=0$. As long as $P_r(r=0)$ is negative and the boundary condition $\phi'(0) = 0$ is imposed, the energy density at the center, $\rho(r=0) = -P_r(r=0)$, is positive. Therefore, the no-go theorem in k-essence theories---which forbids the SSS configuration with a positive-definite energy density---can be circumvented in ESM theories by introducing the $F$-dependence in 
${\cal L}$.

Using the property 
$P_{t}=-P_{r}$, one can 
express Eq.~(\ref{continuity}) in the form 
\be
\hat{P}_r'+\frac{f'}{2f} 
\left( \hat{\rho} 
+ \hat{P}_r \right)=0\,,
\label{balance}
\ee
where 
\ba
\hat{P}_r &\equiv& 
P_r+\int^r 
\frac{4}{\tilde{r}} 
P_r (\tilde{r})
{\rm d} \tilde{r}\,,
\label{hP0}\\
\hat{\rho}  &\equiv& 
\rho-\int^r 
\frac{4}{\tilde{r}} 
P_r (\tilde{r})
{\rm d} \tilde{r}\,.
\label{hP}
\ea
Equation~(\ref{balance}) is analogous to the balance equation for a perfect fluid, with an effective energy density $\hat{\rho}$ and an effective radial pressure $\hat{P}_r$.
Provided that $\hat{P}_r' < 0$, the pressure gradient $\hat{P}_r'$ can counterbalance the gravitational force  $f'/(2f)\,(\hat{\rho} + \hat{P}_r)$. In the following sections, we explicitly show that this behavior indeed occurs in ESM theories.

\section{Purely electric objects} 
\label{electricsec}

In this section, we focus on the purely electric case, characterized by
\be
q_E \neq 0\,,\qquad 
q_M=0\,.
\ee
Setting $q_M = 0$ in Eq.~(\ref{A0con}), the nonvanishing solution for the electric field is given by 
\ba
A_0' &=& \Mpl^2 [2N^2 (r^2 h''-2h+2)
+r N(2r h N''+3r h' N'
\nonumber \\
&&+2 h N')-r^2 h N'^2]/(4N^{3/2}q_E)\,.
\label{rA0}
\ea
Using Eq.~(\ref{back3}), 
the coupling $\mu$ can be 
written as
\be
\mu=\frac{q_E \sqrt{N}}
{r^2 A_0'}\,.
\label{mu}
\ee
Then, the energy density associated 
with the electric field, 
$\rho_{E}=\mu A_0'^2/(2N)$, 
is expressed as 
\be
\rho_{E}=\frac{q_E A_0'}
{2r^2 \sqrt{N}}\,.
\label{rhoA0}
\ee
For the scalar-field derivative appearing in Eq.~(\ref{back2}), one can choose the positive branch of $\phi'$ without loss of generality, yielding
\be
\phi'=\Mpl \sqrt{\frac{N'}{rN}}\,.
\label{rphi}
\ee
The energy density associated with the scalar field, $\rho_{\phi}=h \phi'^2/2$, 
is given by 
\be
\rho_{\phi}=
\frac{\Mpl^2 h N'}{2rN}\,.
\label{rhophi}
\ee
The sum of Eqs.~(\ref{rhoA0}) and (\ref{rhophi}) gives the total energy density, $\rho=\rho_{E}+\rho_{\phi}$.
Using Eq.~(\ref{back1}), the electric 
field can be written as 
\be
A_0'=-\frac{\Mpl^2 
[2(rh'+h-1)N+rhN']}{q_E \sqrt{N}}\,.
\label{rA02}
\ee
Substituting Eq.~(\ref{rA02}) into Eq.~(\ref{mu}), we find 
\be
\mu=-\frac{q_E^2 N}
{\Mpl^2 r^2 [2(rh'+h-1)N+rhN']}\,.
\label{mu2}
\ee
Using Eqs.~(\ref{rA0}) and (\ref{rA02}), we obtain a differential equation relating the two metric functions:
\ba
& &
2 \left( r^2 h''+4r h'+2h-2 
\right)N^2-r^2 h N'^2 
\nonumber \\
& &
+3r( rh'+2h) NN'
+2r^2 h N N''=0\,.
\label{hNeq}
\ea
The solution to Eq.~(\ref{hNeq}), consistent with the regularity condition of 
$h(r)$ in Eq.~(\ref{hr}), is
\be
h(r)=\frac{2 \int_0^r \sqrt{N(r_1)} 
(\int_0^{r_1} \sqrt{N(r_2)} 
{\rm d}r_2) {\rm d}r_1}
{r^2 N(r)}\,, 
\label{hr2}
\ee
where two integration constants must vanish to ensure regularity at $r = 0$.
Since $N(r) > 0$, the function $h(r)$ remains positive for all $r > 0$, implying the absence of a horizon.
In other words, ESM theories do not admit BH solutions with regular centers.

For a given functional form of $N(r)$, the function $h(r)$ can be obtained by integrating 
Eq.~(\ref{hr2}).
Then, from Eqs.~(\ref{rphi}), 
(\ref{rA02}), and (\ref{mu2}), 
one can determine $\phi'$, $A_0'$, 
and $\mu$, as functions of $r$. 
We note that $A_0'$ and $\mu$ contain 
the electric charge $q_E$, so that 
each of them depends on $q_E$. 
However, the energy density 
of the electric field, 
\be
\rho_{E}=-\frac{\Mpl^2
[2(rh'+h-1)N+rhN']}{2r^2 N}\,,
\label{rhoA02}
\ee
does not explicitly depend on $q_E$. The same property also holds for $\rho_\phi$, $\rho$, $P_r$, and $P_t$. 
This implies that, for a given function $N(r)$, the quantities associated with the energy density and 
pressures---such as the mass and radius of the object---are independent of the choice of $q_E$.

Under the expansion of $N(r)$ near $r=0$ given in Eq.~(\ref{Nr}), 
the leading-order contribution to Eq.~(\ref{rphi}) is  
$\phi'(r)=\Mpl \sqrt{2N_2/N_0}+{\cal O}(r)$, which is nonvanishing at $r=0$. 
To avoid the formation of a cusp at $r=0$, which would violate the assumption of spherical symmetry, we require $N_2 = 0$. 
By examining the energy density, 
we further find that, at leading order, 
$\rho= 3\Mpl^2 N_3\,r/N_0+\mathcal{O}(r^2)$. 
Once again, to avoid a cusp at the origin 
in the energy density, we require that $N_3=0$. 
An example of a model satisfying 
$N_2 =0=N_3$ is \cite{DeFelice:2024ops}
\be
N=\left( \frac{r^4
+\sqrt{N_0}\,r_0^4}
{r^4+r_0^4} \right)^2
=\left( \frac{x^4+\sqrt{N_0}}
{x^4+1} \right)^2\,,
\label{Nexample}
\ee
where $N_0$ and $r_0$ are positive constants, and 
\be
x \equiv \frac{r}{r_0}\,.
\ee
In this case, we have 
$\phi'(r)=2\Mpl \sqrt{N_4/N_0}\,r
+{\cal O}(r^2)$ near $r=0$, 
with $N_4=2\sqrt{N_0}(1-\sqrt{N_0})/r_0^4$, and hence $\phi'(0)=0$. 
Since $N_4$ must be positive for the existence of the leading-order solution to $\phi'(r)$ 
around $r=0$, $N_0$ should be in the range
\be
0<N_0<1\,.
\ee
Note that $N(r)$ approaches 1 as $r \to \infty$, so that $f(r) \to h(r)$ in this limit.
For $N_0 = 1$, we have $N(r) = 1$ for any $r$, and hence $h(r) = 1$ from 
Eq.~\eqref{hr2}. This corresponds to the Minkowski spacetime, in which 
$\phi'$ and $A_0'$ vanish; see Eqs.~\eqref{rphi} and \eqref{rA02}. 
In the opposite limit $N_0 \to 0$, the leading-order term of the Ricci scalar near the center, 
$R = 
8(1-\sqrt{N_0})r^2/(\sqrt{N_0} r_0^4)$, diverges, indicating the presence of a spacetime singularity. 
Therefore, we focus on the range $0 < N_0 < 1$.

For the choice \eqref{Nexample}, 
we can integrate Eq.~\eqref{rphi} 
to find
\begin{equation}
\phi=\phi_0
+\Mpl \lambda\,F\!\left(\arcsin\!
\left(\frac{x^2}{\sqrt{x^4+1}} 
\right), -\frac{\lambda^2}{2} \right)\,,
\label{phiso}
\end{equation}
where $F$ is the elliptic integral of the first kind, and 
\be
\lambda \equiv 
\sqrt{\frac{2(1-\sqrt{N_0})}{\sqrt{N_0}}}\,,
\ee
so that $\lambda>0$. Here, $\phi_0$ is an integration constant corresponding to the field value at $r = 0$. 
For a given value of $\lambda$, 
$N_0$ is fixed to be 
$N_0 = 4/(\lambda^2+2)^2$.
Since 
$\phi'(r) > 0$, the scalar field increases monotonically and reaches its maximum value at spatial infinity:
\be
\phi_{\infty}=\phi_0+
\Mpl \lambda\,K(
-\lambda^{2}/2)\,,
\label{phiinf}
\ee
where $K$ denotes the complete elliptic integral of the first kind.
The scalar field is in the 
range $0\leq\phi-\phi_0<\Mpl \lambda\,
K(-\lambda^{2}/2)$.
The relation between $\phi$ and $x = r/r_0$ in Eq.~(\ref{phiso}) can be inverted to give
\be
x=\sqrt{{\rm sc}\!\left[\frac{\phi
-\phi_0}{\lambda\Mpl},-\frac{\lambda^2}2 \right]}\,,
\label{xmu}
\ee
where ${\rm sc}$ denotes the Jacobi elliptic function. 
As we will see below, the relation (\ref{xmu}) allows us to express $\mu$ explicitly as a function of $\phi$, i.e., $\mu = \mu[x(\phi)]$. 

One can perform one of the integrals in 
Eq.~(\ref{hr2}), as
\be
\int_0^r\!\sqrt{N(r_1)}\,{\rm d}r_1
=r_0 \left( x-\frac{1-\sqrt{N_0}}
{2\sqrt{2}} \Delta_1 \right)\,,
\ee
where
\ba
\Delta_1 &\equiv&
\arctan \left(\sqrt{2}x+1 \right)
+\arctan \left(\sqrt{2}x-1 \right) 
\nonumber\\
& &+{\rm arctanh} \left[
\sqrt{2}x/(x^2+1) \right]\,.
\ea
After performing the second integral 
in Eq.~(\ref{hr2}), it follows that 
\ba
h&=&\frac{\left(x^4+1\right)^2}
{8x^2 \left( x^4+\sqrt{N_0} 
\right)^2}
\left[ 4\sqrt{2} \Delta _1
\bigl(\sqrt{N_0}-1\bigr)x
\right.\nonumber\\
&&+\left.\Delta_1^2 \bigl(\sqrt{N_0}-1\bigr)^2
+8 x^2 \right] \,.
\label{hana}
\ea
As we see from Eqs.~(\ref{Nexample}) and (\ref{hana}), both $N$ and $h$ depend only on $x$ and $N_0$.
After normalizing $\rho_{\phi}$ and $\rho_{E}$ as
$\tilde{\rho}_{\phi}=
\rho_{\phi} r_0^2 / \Mpl^2$ and
$\tilde{\rho}_{E}=
\rho_{E} r_0^2 / \Mpl^2$ in Eqs.~(\ref{rhophi}) and (\ref{rhoA02}), respectively,
the same property holds for $\tilde{\rho}_{\phi}$ and $\tilde{\rho}_{E}$.
Therefore, the normalized energy density $\tilde{\rho} = \rho\, r_0^2 / \Mpl^2$ and the normalized pressures 
$\tilde{P}_r = P_r\, r_0^2 / \Mpl^2$ and $\tilde{P}_t = P_t\, r_0^2 / \Mpl^2$ 
do not explicitly depend on either $r_0$ or $q_E$. 
This implies that the SSS configuration can be realized independently of the values of $r_0$ and $q_E$, as we will confirm in the following. 

By substituting Eqs.~(\ref{Nexample}), (\ref{hana}), and their 
$r$ derivatives into 
Eqs.~(\ref{rphi}), (\ref{rA02}), and (\ref{mu2}), 
we can analytically obtain $\phi'(r)$, $A_0'(r)$, and $\mu$ as explicit functions 
of $r$. It should be noted that both $A_0'$ and $\mu$ depend on the electric 
charge $q_E$. 
Introducing the rescaled 
dimensionless charge, 
\be
\tilde{q}_E \equiv \frac{q_E}
{\Mpl r_0}\,,
\ee
one can express $\mu$ as
\be
\mu=\tilde{q}_E^2\, 
{\cal F}(x,N_0)\,,
\ee
where 
\be
{\cal F}(x,N_0)=
-\frac{N}{x^2[ 2\{ xh'(x)
+h-1\} N+xhN'(x)]}\,.
\label{calF}
\ee
Here, a prime denotes differentiation with respect to $x$. For given values of 
$\tilde{q}_E$ and $N_0$, 
we know $\mu$ as a function of 
$x=x(\phi)$. 
In other words, we have
\begin{equation}
\mu(\phi,\tilde{q}_E^2,\lambda)=\tilde{q}_E^2\,
{\cal F}\bigl[ x=x(\phi,\lambda),
N_0=N_0(\lambda)\bigr]\,.
\label{muphi}
\end{equation}
The function (\ref{muphi}) is defined in the interval 
\be
0 \leq \phi-\phi_0
< \Mpl \lambda\, 
K(-\lambda^{2}/2)\,.
\ee
Since the maximum field value is given by 
$\phi_{\infty}
=\phi_0+\Mpl \lambda 
K(-\lambda^{2}/2)$, 
it is determined once $\phi_0$ and $\lambda$ are specified.

For a fixed value of $\tilde{q}_E$ set by the theory, the corresponding electric charge $q_E$ is determined by the free parameter $r_0$. 
In the above relation, $\phi$ denotes a field, whereas $\lambda$ and $\tilde{q}_E$ are universal constants. Once the theory is specified, the values of $\lambda$ and $\tilde{q}_E$ are determined.
This determines certain properties of the compact object under study, such as the value of $N$ at its center.
However, this does not single out a unique configuration for such compact objects, since $r_0$ remains a free parameter that must be specified. Only after fixing $r_0$ can one determine the ADM mass of the object, its charge, and all other physical properties.
From a phenomenological point of view, once observations determine one of these quantities, say the ADM mass, all other physical properties of the object should then be uniquely determined. For instance, as already mentioned, $q_E / r_0 = \tilde{q}_E \Mpl$, which is a universal constant.

\subsection{Small-distance behavior}
\label{smallsec}

Let us examine the behavior of solutions around the center of electrically charged objects for the choice (\ref{Nexample}). 
Using Eq.~(\ref{hana}) with the function (\ref{Nexample}), the metric functions near $r=0$ can be expanded as
\ba
h &=& 1-\frac{4\lambda^2}{5}x^4
+{\cal O} (x^8)\,,
\label{hsmall} \\
f &=& N_0 \left[ 1+
\frac{\lambda^2}{5}x^4 
+{\cal O} (x^8) \right]\,.
\label{fr=0}
\ea
Therefore, $h$ is a decreasing 
function of $x=r/r_0$ around the center, while $f$ increases with $x$.
Using Eqs.~(\ref{rphi}), 
(\ref{rA02}), and (\ref{mu2}), 
we also find 
the following properties 
near $r=0$:
\ba
\frac{{\rm d}\phi}
{{\rm d}x} &=& 2\Mpl \lambda x
+{\cal O}(x^5)\,,\label{rphiD}\\
\frac{{\rm d}A_0}{{\rm d}x}
&=&\frac{8\Mpl \lambda^2}{\tilde{q}_E 
(\lambda^2+2)}x^4
+{\cal O}(x^8)\,,
\label{rA0D}\\
\mu &=& \frac{\tilde{q}_E^2}
{4\lambda^2 x^6}+{\cal O}(x^{-2})\,.
\label{muf}
\ea
One can integrate Eq.~(\ref{rphiD}) to give
\be
\phi=\phi_0+\Mpl \lambda
x^2+{\cal O}(x^6)\,.
\label{phiso2}
\ee
Although $\mu$ diverges at $r=0$, the product $\mu F$ in the Lagrangian (\ref{model}) remains finite. In fact, since $F \propto r^8$, we have $\mu F \propto r^2$, which vanishes in the limit $r \to 0$. From Eqs.~(\ref{muf}) and (\ref{phiso2}), the leading-order term of $\mu$ has the following $\phi$ dependence: 
\be
\mu (\phi) \simeq 
\frac{{\tilde q}_E^2 \Mpl^3 \lambda}
{4(\phi-\phi_0)^3}\,,
\ee
which is positive.
Therefore, the coupling must behave as $\mu(\phi) \propto (\phi - \phi_0)^{-3}$ to realize the form of 
$N(r)$ in Eq.~(\ref{Nexample}) near the center. 
The divergence of $\mu(\phi)$ at $r=0$ is consistent 
with the existence of regular compact objects 
in ESM theories, as demonstrated 
in Ref.~\cite{Herdeiro:2019iwl}.

Expanding the energy density $\rho$ and the pressures $P_r$ and $P_{t}$ around $r=0$, it follows that 
\ba
\hspace{-0.7cm}
\rho &=& 
\frac{4 \Mpl^2 \lambda^2}{r_0^2}x^2+{\cal O}(x^6)\,,\label{rhor=0}\\
\hspace{-0.7cm}
P_r &=& -P_{t}
=-\frac{4\Mpl^2 \lambda^4}{25r_0^2}x^6+{\cal O}(x^{10})\,.
\label{Pr=0}
\ea
In the limit $r \to 0$, 
all of $\rho$, 
$P_r$, and $P_t$ vanish.
This feature can be 
primarily attributed 
to the requirement $N_2 = 0$,
introduced to ensure that $\phi'(r)$ 
vanishes at the origin. 
Unlike the usual case in which 
$h(r)$ is expanded as 
$h(r)=1+h_2 r^2+\cdots$ around $r=0$, the next-to-leading order term of $h(r)$ is proportional to $r^4$. 

Defining the equations of state 
as $w_r = P_r/\rho$ and 
$w_t = P_t/\rho$ for the radial and angular directions, respectively, 
their leading-order behavior near the center is given by
\be
w_r=-w_{t}=
-\frac{\lambda^2}{25}x^4\,.
\ee
In the limit $x \to 0$, we have $|w_r| = w_t \ll 1$, 
indicating that the matter sector is in a 
nonrelativistic regime 
near the center.
This behavior is markedly different from that in NSs 
\cite{Burgio:2021vgk} and gravastars with a de Sitter condensate core \cite{Mazur:2001fv}.
As $r$ approaches $r_0$, $|w_r|$ and $w_t$ increase, potentially entering a relativistic regime characterized by $|w_r| = w_t \gtrsim \mathcal{O}(0.1)$. 
For smaller values of $N_0$ (i.e., larger $\lambda$), the magnitudes of $|w_r|$ and $w_t$ near $r = r_0$ tend to be larger.

The effective radial pressure $\hat{P}_r$ and energy density 
$\hat{\rho}$, defined in 
Eqs.~(\ref{hP0}) and (\ref{hP}), take the form 
\ba
\hat{P}_r 
&=& {\cal C}_0
-\frac{4\Mpl^2 \lambda^4}
{15 r_0^2}x^6
+{\cal O} (x^{10})\,,
\label{hPr=0}\\
\hat{\rho} 
&=& -{\cal C}_0
+\frac{4\Mpl^2 \lambda^2}{r_0^2}
x^2+{\cal O}(x^{6})\,,
\label{hrhor=0}
\ea
where ${\cal C}_0$ is an integration constant.
Using Eqs.~(\ref{fr=0}), (\ref{hPr=0}), and 
(\ref{hrhor=0}), we find 
that the balance relation 
${\rm d}\hat{P}_r/{\rm d}x
=-8\Mpl^2\lambda^4 x^5/(5r_0^2)
=-(\hat{\rho}+\hat{P}_r)
({\rm d}f/{\rm d}x)/(2f)$, 
which corresponds to 
Eq.~(\ref{balance}), 
is indeed satisfied.

\subsection{Large-distance behavior}
\label{largesec}

Let us proceed to study the behavior of solutions in the large-distance regime characterized by $r \gg r_0$.
From Eqs.~(\ref{Nexample}) and (\ref{hana}), the metric functions can be expanded as 
\ba
h &=& 1-\frac{\sqrt{2}\pi 
\lambda^2}{2(\lambda^2+2)x}
+\frac{\pi^2 \lambda^4}{8(\lambda^2+2)^2 x^2} 
\nonumber \\
& &+\frac{8\lambda^2}{3(\lambda^2+2)x^4}
+{\cal O}(x^{-5})\,,\label{hinf}\\
f &=& 1-\frac{\sqrt{2}\pi \lambda^2}
{2(\lambda^2+2)x}
+\frac{\pi^2 \lambda^4}{8(\lambda^2+2)^2 x^2} 
\nonumber \\
& &+\frac{2\lambda^2}{3(\lambda^2+2)x^4}
+{\cal O}(x^{-5})\,.\label{finf}
\ea
The difference between $h$ and $f$ appears at order $x^{-4}$. Substituting Eq.~(\ref{finf}) into the definition of the ADM mass $M$ in Eq.~(\ref{ADM}), 
we find
\be
M=\frac{2\sqrt{2}\pi^2
\lambda^2}
{\lambda^2+2}M_0\,,
\label{MADM}
\ee
where 
\be
M_0 \equiv \Mpl^2 r_0
=2.7 \times 10^{-2} M_{\odot} 
\left( \frac{r_0}
{1~{\rm km}} \right)\,,
\label{M0}
\ee
and $M_{\odot} = 1.989 \times 10^{30}\,{\rm kg}$ denotes the solar mass. 
In Sec.~\ref{mrasec}, 
we will study the 
mass-radius relation of the compact object in detail.

From Eqs.~(\ref{rphi}), (\ref{rA02}), and (\ref{mu2}), 
the large-distance behavior of 
${\rm d}\phi/{\rm d}x$, 
${\rm d}A_0/{\rm d}x$, and 
$\mu$ is given, 
respectively, by
\ba
\hspace{-0.5cm}
\frac{{\rm d}\phi}
{{\rm d}x} 
&=& \frac{2\sqrt{2}\Mpl 
\lambda}
{\sqrt{\lambda^2+2}}
\frac{1}{x^3}+{\cal O} (x^{-7})\,,
\label{rphiinf}
\\
\hspace{-0.5cm}
\frac{{\rm d}A_0}{{\rm d}x} &=& 
\frac{\pi^2\Mpl \lambda^4}{4\tilde{q}_E 
(\lambda^2+2)^2}
\frac{1}{x^2}+{\cal O} (x^{-4})\,,
\label{A0inf}
\\
\hspace{-0.5cm} 
\mu &=& 
\frac{4 \tilde{q}_E^2 
(\lambda^2+2)^2}{\pi^2 
\lambda^4} 
\biggl[ 1-\frac{32(\lambda^2+2)}
{\pi^2 \lambda^2}\frac{1}{x^2} \nonumber \\
\hspace{-0.5cm} 
& &\qquad\qquad\qquad\quad
+{\cal O}(x^{-3})
\biggr],
\label{muinf}
\ea
where we have taken into account the next-to-leading order term in $\mu$.
One can integrate 
Eq.~(\ref{rphiinf}) to give
\be
\phi=\phi_{\infty}
-\frac{\sqrt{2} \Mpl \lambda}
{\sqrt{\lambda^2+2}} 
\frac{1}{x^2}+{\cal O} (x^{-6})\,.
\label{phiinfD}
\ee
From Eqs.~(\ref{muinf}) and (\ref{phiinfD}), the coupling $\mu(\phi)$ has the following dependence:
\be
\mu(\phi) \simeq \frac{4\tilde{q}_E^2 (\lambda^2+2)^2}{\pi^2 \lambda^4} \left[ 1+\frac{16\sqrt{2}(\lambda^2+2)^{3/2}
(\phi-\phi_\infty)}{\pi^2\Mpl \lambda^3}
\right].
\label{muasy}
\ee
As $\phi$ increases toward its asymptotic value $\phi_{\infty}$, the coupling $\mu(\phi)$ approaches the following constant:
\be
\mu_{\infty}=
\frac{4\tilde{q}_E^2 (\lambda^2+2)^2}{\pi^2 \lambda^4}
=\frac{4\tilde{q}_E^2}
{\pi^2 (1-\sqrt{N_0})^2}\,, 
\ee
which is positive. 
In the region $r \ll r_0$, we recall that $\mu(\phi)$ decreases as a function of $\phi$, i.e., $\mu(\phi) \propto (\phi - \phi_0)^{-3}$.
To match this small-distance solution with the large-distance behavior given in Eq.~(\ref{muasy}), $\mu(\phi)$ must attain a minimum value 
$\mu_{\rm min}$ 
at an intermediate field value.
To ensure the no-ghost condition $\mu(\phi) > 0$ 
for any value of $\phi$, we require $\mu_{\rm min} > 0$.
In Sec.~\ref{intersec}, we will confirm that this condition is indeed satisfied.

Up to the order of $r^{-2}$, 
the metric functions $h$ and $f$ in Eqs.~(\ref{hinf}) and (\ref{finf}) 
can be written as 
$h=f=1-M/(4\pi \Mpl^2 r)
+q_E^2/(2\Mpl^2 \mu_{\infty} r^2)$. From Eq.~(\ref{A0inf}), the electric field behaves as 
$A_0'(r) = q_E/(\mu_{\infty} r^2)$ at leading order, so that the SSS object has an effective electric charge 
$q_E / \mu_{\infty}$.  
From Eq.~(\ref{rphiinf}), the derivative of the scalar field, $\phi'(r)$, decreases faster than $r^{-2}$ at large distances. Therefore, the compact object does not possess a scalar charge.

We also find that, at large distances, the solutions for $\rho$, $P_r$, and $P_t$ are given by
\ba
\rho &=& \frac{\pi^2 \Mpl^2 
\lambda^4}
{8(\lambda^2+2)^2 r_0^2} 
\frac{1}{x^4}+{\cal O} (x^{-6})\,,\\
P_r &=& -P_t=
-\frac{\pi^2 \Mpl^2 \lambda^4}
{8(\lambda^2+2)^2 r_0^2} 
\frac{1}{x^4}+{\cal O}(x^{-7})\,,
\label{Prinf}
\ea
both of which are independent of $q_E$.
The corresponding equations 
of state, 
$w_r = P_r/\rho$ and $w_t = P_t/\rho$, 
at leading order, are
\be
w_r=-1\,,\qquad 
w_t=1\,.
\ee
Due to the property $w_r=-1$, 
the gravitational force 
$f'/(2f) (\hat{\rho}+\hat{P}_r)$ 
in Eq.~(\ref{balance}), 
which is equivalent to 
$f'/(2f) (\rho+P_r)$, 
vanishes.
Using the leading-order solution for $P_r$ in Eq.~(\ref{Prinf}), the effective pressure gradient $\hat{P}_r'$ also vanishes, 
so that the balance relation (\ref{balance}) is satisfied at large distances.

\subsection{Solutions covering the intermediate-distance regime}
\label{intersec}

In Secs.~\ref{smallsec} and \ref{largesec}, we studied the properties of the solutions near $r=0$ and at large distances. In this section, we investigate the behavior of solutions at arbitrary distances $r$ to understand how the two asymptotic solutions are connected 
in the intermediate region.
Since we have an analytic solution for $h(x)$ as well 
as for $N(x)$, all other quantities, such as 
${\rm d}\phi/{\rm d}x$, $\phi-\phi_0$, ${\rm d}A_0/{\rm d}x$, $\mu$, $\rho_{\phi}$, $\rho_{E}$, $\rho$, and 
$P_r=-P_t$, are determined 
for given values of $N_0$ 
and $\tilde{q}_E$.
We also numerically integrate the differential equation~(\ref{hNeq}) for $h(x)$, starting from the vicinity 
of $x = 0$.  
We confirm that the numerical solution for $h(x)$ is in excellent agreement with the analytic one.

\begin{figure}[ht]
\begin{center}
\includegraphics[height=3.3in,width=3.4in]{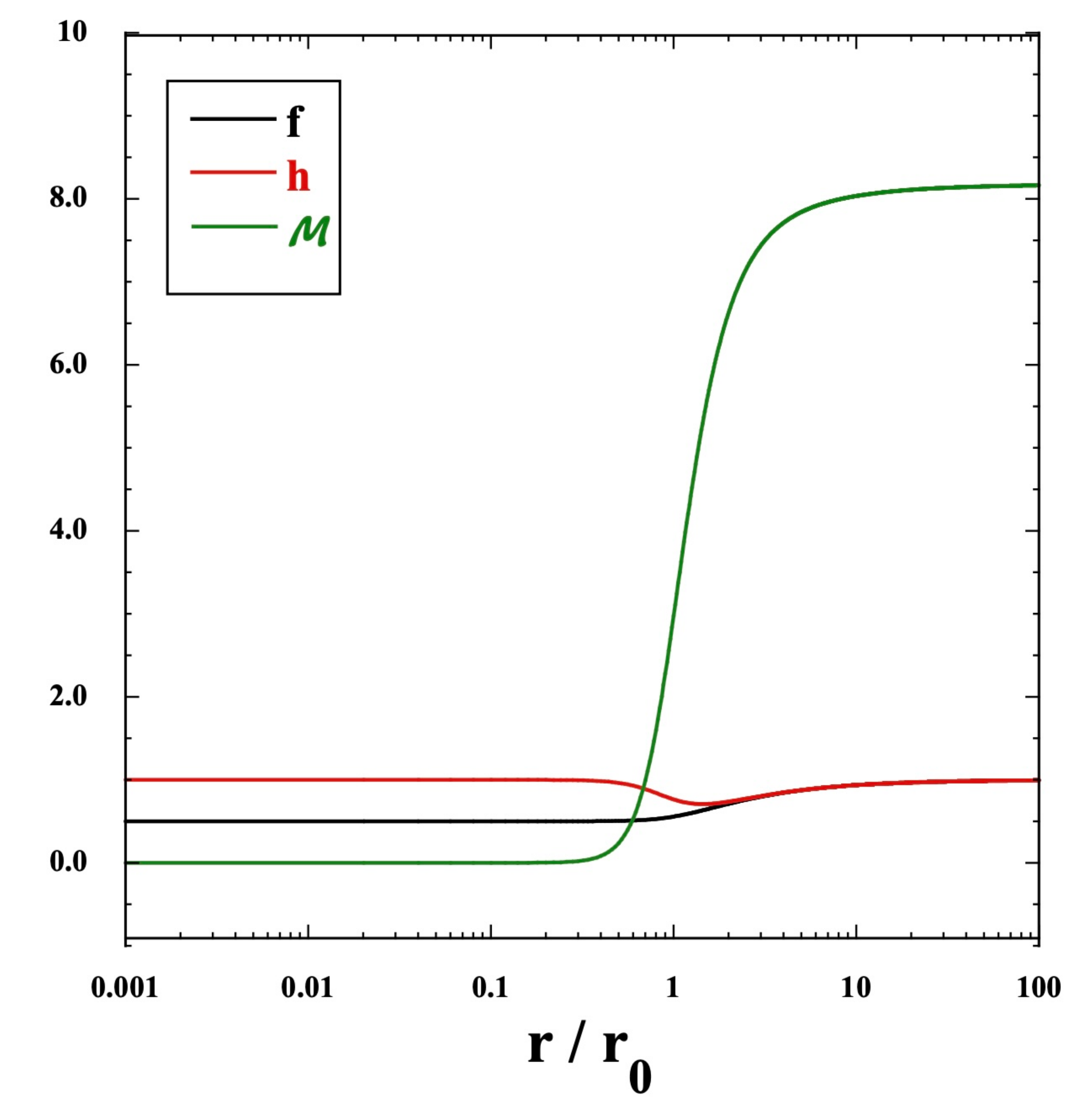}
\end{center}
\caption{Metric components $f$ and $h$ 
as functions of $x=r/r_0$ for $N_0=0.5$.
We also show the mass function ${\cal M}$, 
which is normalized by 
$M_0=\Mpl^2 r_0$.
\label{figmet}}
\end{figure}

In Fig.~\ref{figmet}, we plot $f(x)$, $h(x)$, and ${\cal M}(x)$ as functions of 
$x = r / r_0$ for 
$N_0=0.5$. These quantities are independent of the choice of $\tilde{q}_E$. 
Near the center, $h(x)$ decreases from 1 as $x$ increases.
After reaching a minimum value $h \simeq 0.71$ at 
$x \simeq 1.45$, 
$h(x)$ starts to increase toward its asymptotic value of 1. Since $h(x)$ does not 
cross 0, the SSS object 
does not possess a horizon. 
Indeed, as shown in Eq.~(\ref{hr2}), this property holds for any value of $N_0$ in the range 
$0 < N_0 < 1$.
As $x$ increases, the mass function ${\cal M}(x)$ 
approaches a constant value, 
$M = 8.16\,\Mpl^2 r_0$, 
indicating that the object is compact. 
The other metric function, $f(x)$, starts to vary from 
the value $N_0 = 0.5$ at the center and then continuously increases 
toward its asymptotic value of 1. Since the derivative $f'(r)$ is always positive, the gravitational force is attractive, as expected.

\begin{figure}[ht]
\begin{center}
\includegraphics[height=3.3in,width=3.3in]{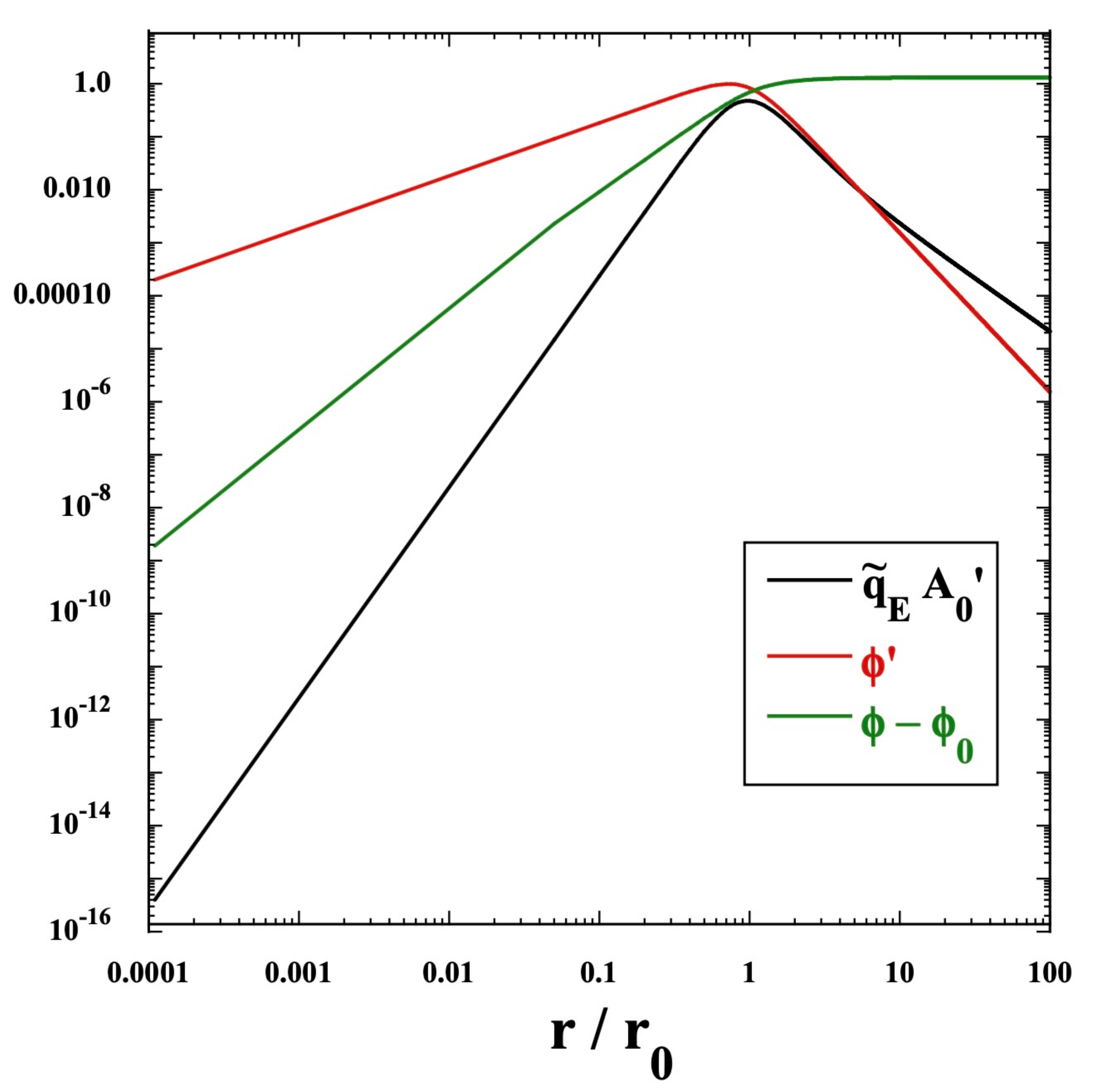}
\end{center}
\caption{Profiles of $\tilde{q}_E A_0'$, $\phi'$, 
and $\phi-\phi_0$ as functions of $x=r/r_0$ 
for $N_0=0.5$. 
Each quantity is normalized by $\Mpl/r_0$, $\Mpl/r_0$, and $\Mpl$, 
respectively.
\label{figAphi}}
\end{figure}

\begin{figure}[ht]
\begin{center}
\includegraphics[height=3.3in,width=3.3in]{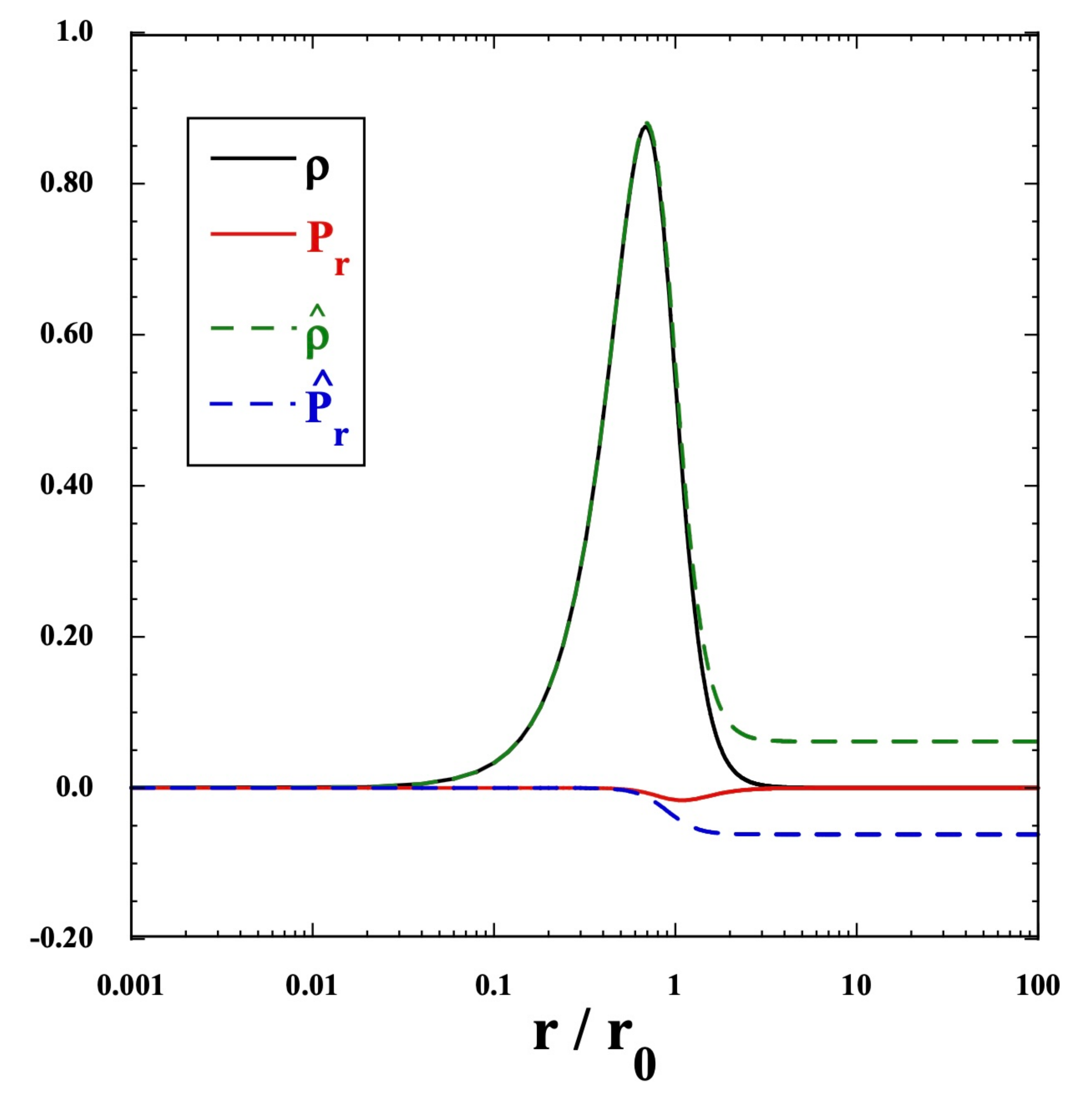}
\end{center}
\caption{Profiles of $\rho$, $P_r$, $\hat{\rho}$, and $\hat{P}_r$ as functions of $r/r_0$ for $N_0=0.5$. 
All of these quantities are 
normalized by $\Mpl^2/r_0^2$.
\label{figrhoP}}
\end{figure}

\begin{figure}[ht]
\begin{center}
\includegraphics[height=3.2in,width=3.4in]{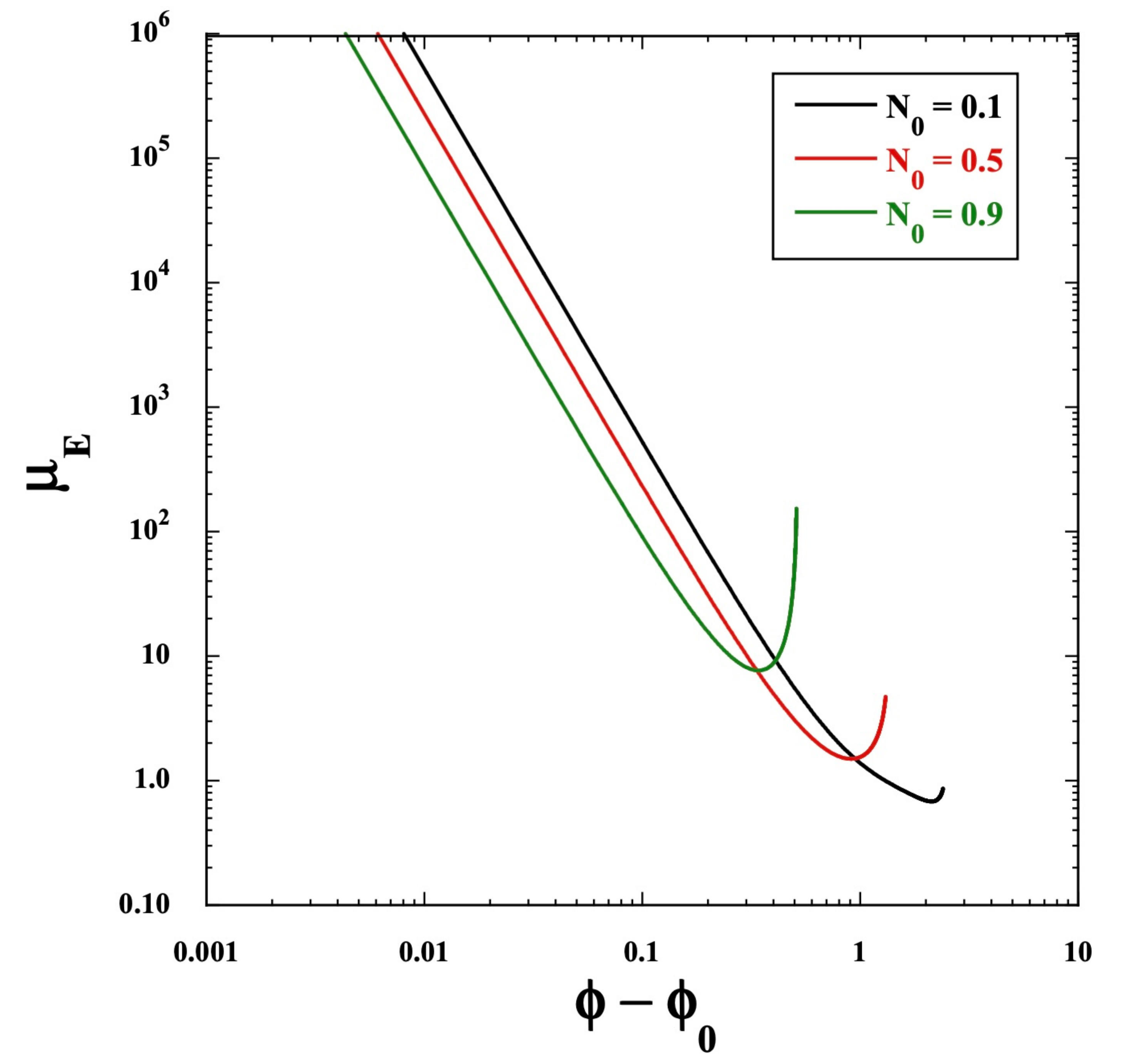}
\end{center}
\caption{The rescaled coupling 
$\mu_E=\tilde{q}_E^{-2}\mu$ as a function of $\phi-\phi_0$ (normalized by $\Mpl$) with three different values of $N_0$.
\label{figmuE}}
\end{figure}

In Fig.~\ref{figAphi}, we show the profiles of $\tilde{q}_E A_0'$, $\phi'$, and $\phi-\phi_0$ 
for $N_0=0.5$. 
As estimated from 
Eqs.~(\ref{rphiD}) and (\ref{rA0D}), 
we find that 
$\phi'(r) \propto r$ 
and $A_0'(r) \propto r^4$ 
near $r=0$.
After reaching their maximum 
values, $\phi'(r)$ and $A_0'(r)$ start 
to decrease around $r=r_0$. 
In the region $r \gg r_0$, the field derivatives behave as $\phi'(r) \propto r^{-3}$ and 
$A_0'(r) \propto r^{-2}$, in agreement with the analytic estimates (\ref{rphiinf}) and (\ref{A0inf}). 
Near $r=0$, the scalar field increases as 
$\phi(r)-\phi_0 \propto r^2$. Asymptotically, $\phi(r)$ approaches its asymptotic value $\phi_{\infty}=\phi_0+
\Mpl \lambda 
K(-\lambda^2/2)$.

In Fig.~\ref{figrhoP}, we plot $\rho$, $P_r$, $\hat{\rho}$, and $\hat{P}_r$ as functions of $r/r_0$ for $N_0 = 0.5$. 
For the computation of $\hat{\rho}$ and $\hat{P}_r$, 
we have set the integration constant ${\cal C}_0$ in the small-distance expansions of Eqs.~(\ref{hPr=0}) and (\ref{hrhor=0}) to 0.
As in the case of the metric functions $f$ and $h$, these quantities do not depend on the value of $\tilde{q}_E$. 
In Fig.~\ref{figrhoP}, we observe that $\rho(r)>0$ and $P_r(r)<0$ at all radii $r$. Near the 
center ($r=0$), we have $\rho(r) \propto r^2$ and $-P_r(r)=P_t(r) \propto r^6$, as estimated from Eqs.~(\ref{rhor=0}) and (\ref{Pr=0}), indicating that the matter fields are nonrelativistic with 
$|w_r| \ll 1$ and $w_t \ll 1$. After $\rho(r)$ and $|P_r(r)|$ reach their 
maximum values around $r=r_0$, they 
begin to decrease and exhibit the asymptotic behavior $\rho(r) \simeq -P_r(r) \propto r^{-4}$ 
in the region $r \gg r_0$.
In Fig.~\ref{figrhoP}, we observe that the effective radial pressure $\hat{P}_r(r)$, defined by Eq.~(\ref{hP0}), decreases monotonically, such that 
$\hat{P}_r'(r)<0$. 
This negative effective pressure gradient counteracts the positive gravitational term, 
$f'/(2f)(\hat{\rho} + \hat{P}_r)=f'/(2f)(\rho + P_r)=
f'/(2f) h \phi'^2$. 
At large distances, as seen in Fig.~\ref{figrhoP}, we have $\hat{P}_r \simeq -\hat{\rho}$, and hence $P_r \simeq -\rho$ with $\hat{P}_r'(r) \simeq 0$. 
We also note that, for physical quantities 
such as $P_r$ and $\rho$, 
we have 
$\lim_{r \to \infty} P_r = 0$ 
and 
$\lim_{r \to \infty} \rho = 0$, 
because we are considering asymptotically flat solutions.
Thus, electric SSS compact objects with peculiar profiles of energy density and pressures exist in ESM theories. 
The key difference from stars composed of perfect fluids is that $\rho$, $P_r$, and $P_t$ all vanish at $r=0$.

In Fig.~\ref{figmuE}, we plot the rescaled coupling 
\be
\mu_E \equiv 
\tilde{q}_E^{-2} \mu
={\cal F}\,,
\label{muE}
\ee
as a function of 
$\phi - \phi_0$ for $N_0 = 0.1$, $0.5$, and $0.9$, 
where ${\cal F}$ is defined in Eq.~(\ref{calF}).
We find that $\mu(\phi) \propto (\phi - \phi_0)^{-3}$ in the region where $\phi$ is close to $\phi_0$, that is, near $r = 0$.
In the region $r \gg r_0$, the scalar field $\phi$ 
and the coupling $\mu$ asymptotically approach  
their respective values, 
$\phi_{\infty} = \phi_0
+ \Mpl \lambda 
K(-\lambda^{2}/2)$ and 
$\mu_{\infty} =4\tilde{q}_E^{2}/[\pi^2 
(1 - \sqrt{N_0})^2]$. 
As seen in Fig.~\ref{figmuE}, 
in the region where $\phi$ approaches $\phi_{\infty}$, 
the coupling $\mu(\phi)$ increases toward its asymptotic value 
$\mu_{\infty}$ at 
$\phi = \phi_{\infty}$.
This behavior is consistent with the analytic solution (\ref{muasy}), 
which indicates that 
$\mu - \mu_{\infty}$ increases linearly with $\phi$.
We also note that the numerical values of $\phi_{\infty}$ and $\mu_{\infty}$ in each case shown in Fig.~\ref{figmuE} 
are in good agreement with the corresponding analytic estimates.
In an intermediate regime of $\phi$, the rescaled coupling $\mu_E(\phi)$ has a minimum value $\mu_{E,{\rm min}}$, which increases with $N_0$.
Since $\mu_{E,{\rm min}}>0$ for values of $N_0$ in the range $0 < N_0 < 1$, the ghost-free condition $\mu(\phi) > 0$ is always satisfied.

\subsection{Mass-radius relation}
\label{mrasec}

The ADM mass $M$ of the 
electric object is analytically known from Eq.~(\ref{MADM}). In terms of $N_0$, it can be expressed as
\be
M=2\sqrt{2} \pi^2 \left( 
1-\sqrt{N_0} \right)M_0\,.
\label{ADM2}
\ee
In the limit $N_0 \to 0$, 
the ADM mass reaches its maximum value
\be
M_{\rm max}=2\sqrt{2}
\pi^2 M_0
=0.75 M_{\odot} \left( 
\frac{r_0}{1~{\rm km}} 
\right)\,.
\label{Mmax2}
\ee
After normalizing the background equations of 
motion (\ref{back1})-(\ref{graeq3}) 
with respect to 
$x = r/r_0$ and 
$\tilde{\rho} =\rho\, r_0^2/\Mpl^2$, etc., they do 
not contain any explicit dependence on $r_0$.
Therefore, $r_0$ is an arbitrary positive constant. 
For $r_0 = 1~{\rm km}$, the maximum mass is $M_{\rm max} = 0.75 M_{\odot}$. 
The ADM mass given in Eq.~(\ref{ADM2}) vanishes in the limit $N_0 \to 1$, 
which corresponds to the Minkowski spacetime where $f(r) = h(r) = 1$ everywhere.

\begin{figure}[ht]
\begin{center}
\includegraphics[height=3.2in,width=3.4in]{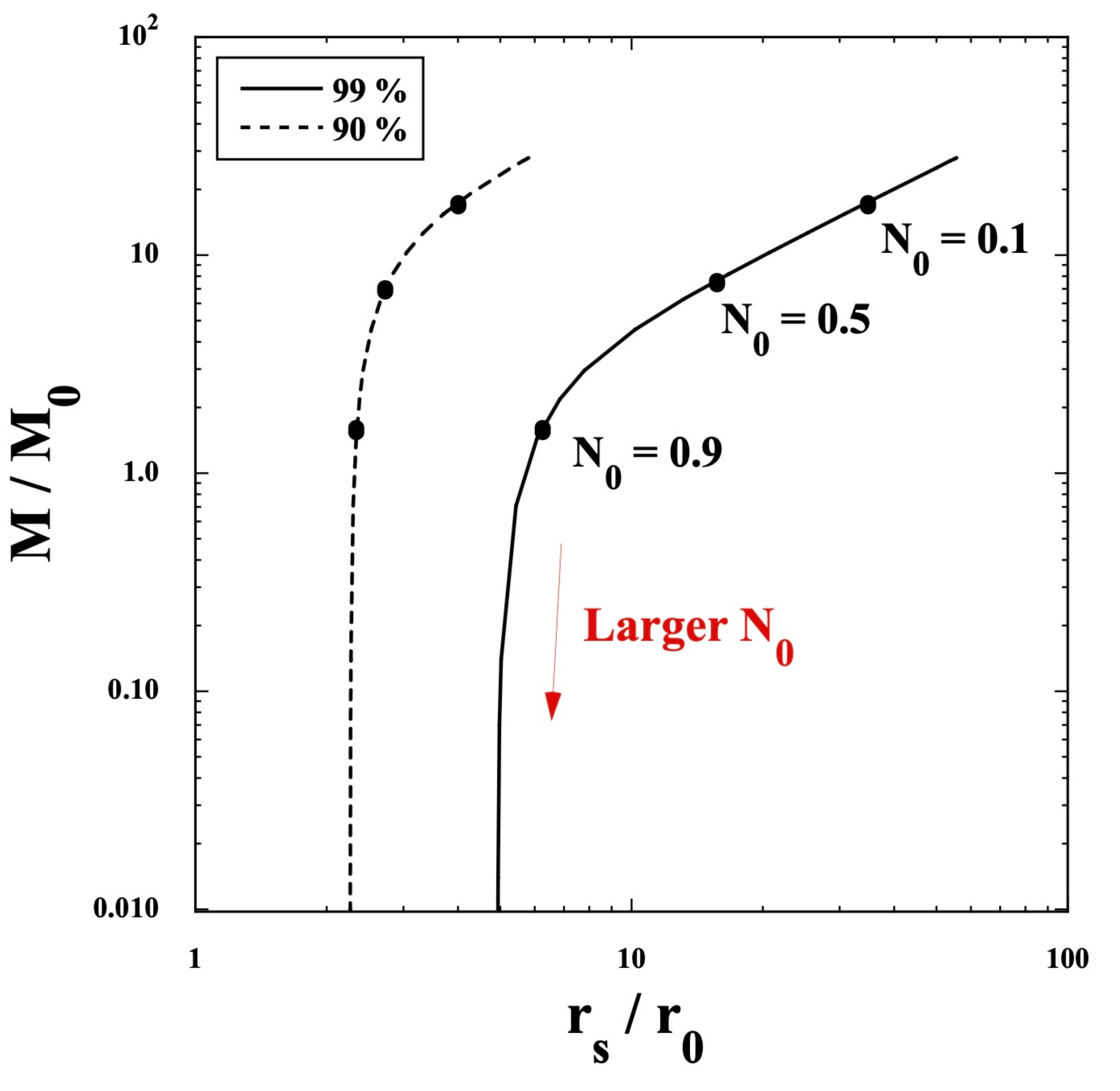}
\end{center}
\caption{The ADM mass $M$ is 
plotted against 
the object's radius $r_s$, 
where $M$ and $r_s$ are normalized by $M_0 = \Mpl^2 r_0$ and $r_0$, respectively. 
The solid and dashed lines represent the cases where the radius is determined by the conditions ${\cal M}(r_s) = 0.99M$ and ${\cal M}(r_s) = 0.90M$, respectively.
The black dots along both the solid and dashed lines correspond to the cases with $N_0 = 0.1$, $0.5$, and $0.9$.
\label{figMr}}
\end{figure}

\begin{figure}[ht]
\begin{center}
\includegraphics[height=3.2in,width=3.4in]{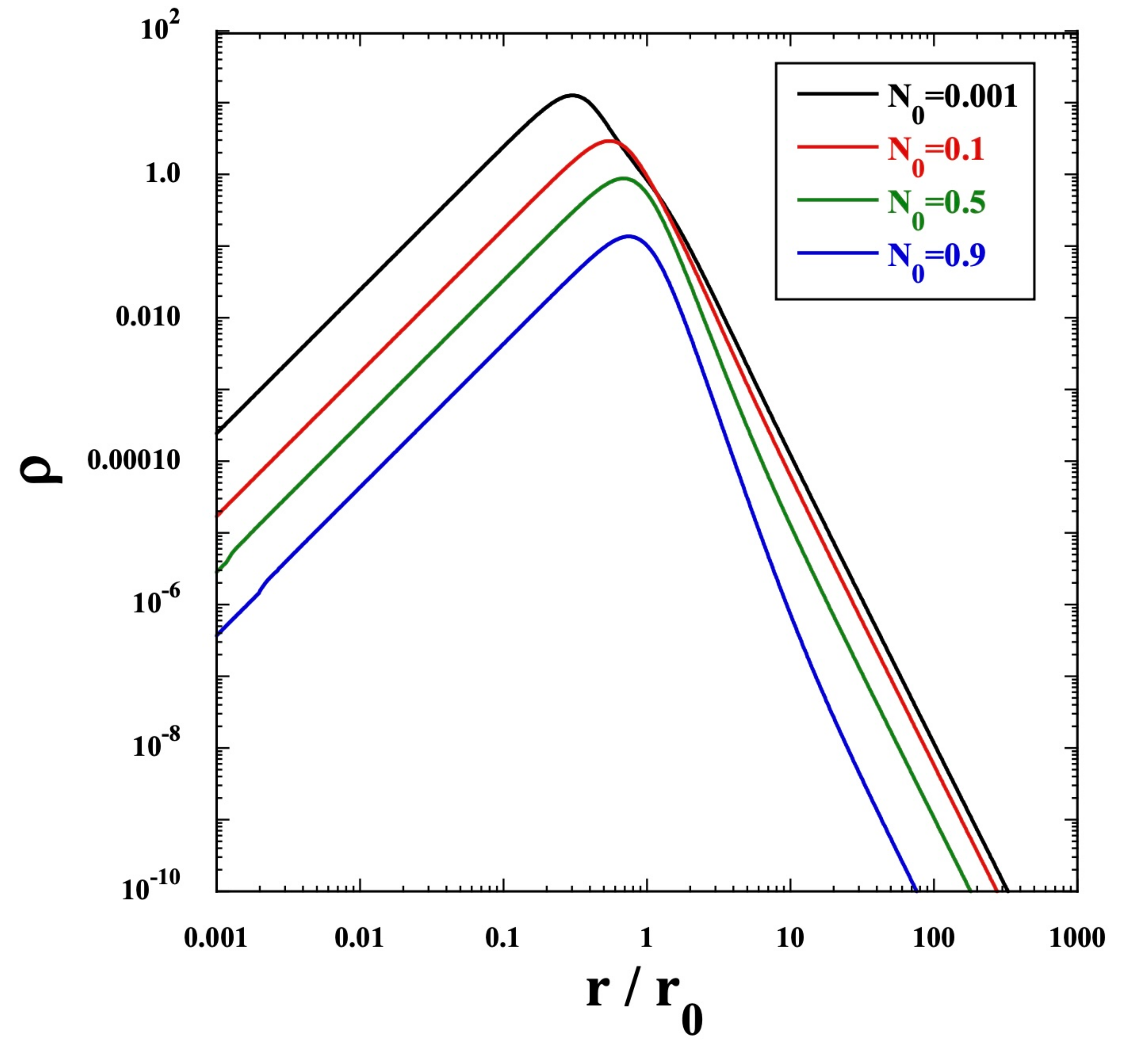}
\end{center}
\caption{The energy density $\rho$ versus $r/r_0$ for 
four different values of $N_0$, where $\rho$ is normalized by $\Mpl^2/r_0^2$. 
\label{figrhoN0}}
\end{figure}

For a star composed of a perfect fluid, the standard definition of the radius $r_s$ is the location where the radial pressure vanishes.
In our case, the effective radial pressure $\hat{P}_r$ always decreases as a function of $r$, but it approaches a constant nonzero value at spatial infinity (see Fig.~\ref{figrhoP} for the 
choice ${\cal C}_0=0$). 
Therefore, instead of using the pressure, we define the radius $r_s$ as the point where the mass function ${\cal M}(r)$ reaches 99\% of the ADM mass $M$, i.e., ${\cal M}(r_s) = 0.99\,M$.\footnote{This implies that the radius corresponds to the value of 
$x_s = r_s/r_0$ at which 
$2x_s [1-h(x_s)]
-\sqrt{2} \pi(1-\sqrt{N_0})
\beta=0$, where $\beta = 0.99$ is a chosen value.}
This percentage can be chosen more flexibly, so we also consider the case in which the radius is determined by the condition ${\cal M}(r_s) 
= 0.90\,M$.
For given values of $N_0$, the radius $r_s$ can be computed using the analytic expression (\ref{hana}) for $h$, together with the definition (\ref{mfunction}) of the mass function. 
To verify the robustness of the results, we also numerically integrate Eq.~(\ref{hNeq}) for $h$ up to a sufficiently large distance, say $r = 10^6\,r_0$, and compute the mass function to identify the radius according to the condition ${\cal M}(r_s) = 0.99\,M$ (and also 
${\cal M}(r_s) = 0.90\,M$).

After obtaining $M$ and $r_s$, we can compute the compactness of the object, which is defined by 
\be
{\cal C} \equiv 
\frac{M}{8 \pi \Mpl^2 r_s}
=\frac{1}{8\pi} \frac{M}
{M_0} \frac{r_0}{r_s}\,.
\ee
The compactness increases with larger $M$ or smaller $r_s$, but it is independent of $r_0$. 
In other words, once a criterion for defining the radius is fixed, compactness becomes a universal property of these objects.

In Fig.~\ref{figMr}, we show the mass-radius relation for values of $N_0$ in the range 
$0<N_0<1$. The solid and dashed lines represent the cases where $r_s$ is determined by the conditions ${\cal M}(r_s)=0.99M$ and ${\cal M}(r_s)=0.90M$, respectively.
As $N_0$ increases from 0, both $M$ and $r_s$ decrease accordingly. To understand this behavior, we plot $\rho(r)$ as a function of $r$ for $N_0=0.001, 0.1, 0.5, 0.9$ in Fig.~\ref{figrhoN0}. 
In the regime $r \ll r_0$, the energy density follows the behavior given by Eq.~(\ref{rhor=0}), such that $\rho(r)$ becomes smaller as $N_0$ increases 
toward 1. For larger $N_0$, the maximum values of $\rho(r)$ reached at $r=r_m$ are also subject to decrease. 
Although $r_m$ increases with larger $N_0$, the radius $r_s$ tends to decrease due to the shift of the $\rho(r)$ curves toward smaller $r$ for $r > r_m$ (see Fig.~\ref{figrhoN0}). 
This explains the reduced 
values of $r_s$ for increasing $N_0$.

As $N_0$ decreases, the maximum energy density $\rho(r_m)$ increases, with $r_m$ shifting toward smaller values. 
The product $\rho(r_m) r_m^2$ multiplied by $r_s$, i.e., 
${\cal M}_0 \equiv \rho(r_m) r_m^2 r_s$, is roughly proportional to the ADM mass $M$ of the object.
As long as $N_0$ is not very close to 0, both $\rho(r_m) r_m^2$ and $r_s$ increase as $N_0$ decreases (see Fig.~\ref{figrhoN0}), 
leading to increases in 
${\cal M}_0$ and $M$.
However, as $N_0$ approaches 0, the increases in $\rho(r_m) r_m^2$ and $r_s$ become saturated, so that ${\cal M}_0$, $M$, and $r_s$ approach constant values. Therefore, for a given $r_0$, we obtain the maximum ADM mass given by Eq.~(\ref{Mmax2}). 
Using the condition 
${\cal M}(r_s)=0.99M$, the 
corresponding maximum 
radius is
\be
(r_s)_{\rm max}
= 55.6\,r_0\,.
\label{rmax}
\ee
For $r_0 = 2$~km, 
we have $(r_s)_{\rm max} = 1.12 \times 10^2$~km and 
$M_{\rm max} = 1.5\,M_{\odot}$.
The compactness corresponding to 
the maximum mass and the 
radius (\ref{rmax}) is 
${\cal C} = 0.02$. 
For $r_0 \gg 1$~km, 
it is possible to realize 
the SSS configuration 
with $M \gg M_{\odot}$ 
and $(r_s)_{\rm max} \gg {\cal O}(10~{\rm km})$. 

As $N_0$ increases from 0 to 1, the ADM mass given by Eq.~\eqref{ADM2} decreases from $M_{\rm max}$ toward 0, while the radius $r_s$ simultaneously becomes smaller than its maximum value given in Eq.~\eqref{rmax}.
As an example, for $N_0 = 0.5$, we obtain 
$M = 8.18\,M_0$ and 
$r_s = 16.7\,r_0$, with a compactness of 
${\cal C} = 0.02$. 
This is much larger than the compactness of white dwarfs, 
${\cal C} = \mathcal{O}(10^{-4})$, 
and that of the Sun, 
${\cal C} = \mathcal{O}(10^{-6})$, but still 
smaller than that of a BH, $\mathcal{C}=1/2$. 
For values of $N_0$ in the range $0 < N_0 
\lesssim 0.7$, 
we find that $M$ is approximately proportional to $r_s$, see Fig.~\ref{figMr}.
Therefore, the compactness remains nearly constant 
(${\cal C} \simeq 0.02$). 
For $N_0 \gtrsim 0.9$, the ADM mass rapidly decreases below $M_0$, resulting a compactness much smaller than ${\cal O}(10^{-2})$.
In the limit $N_0 \to 1$, we have $N(r) \to 1$ at any distance $r$, and hence $h(r) \to 1$ from Eq.~(\ref{hana}).
In this limit, we also find $A_0'(r) \to 0$ and $\phi'(r) \to 0$ from Eqs.~(\ref{rA0}) and (\ref{rphi}), so that the energy density
$\rho(r) = h \phi'^2 / 2 + \mu A_0'^2 / (2N)$
and the ADM mass $M$ both vanish.
Conversely, for $N_0$ close to 0, the variation of $N(r)$ becomes maximal, leading to the largest values of $A_0'(r)$, $\phi'(r)$, and $M$.

The above results for the radius and compactness have been obtained under the condition ${\cal M}(r_s)=0.99M$. 
If we relax this condition to ${\cal M}(r_s)=0.90M$, we find that the radius corresponding to the maximum ADM mass is given by 
$(r_s)_{\rm max}=5.8r_0$. 
This value is smaller by one order of magnitude than that in Eq.~\eqref{rmax}, so that the compactness increases to 
${\cal C}=0.19$. 
For $N_0 = 0.7$, we obtain $r_s = 2.53r_0$ and $M = 4.56M_0$, giving a compactness of ${\cal C} = 0.07$. 
For $N_0 \gtrsim 0.7$, the compactness rapidly drops below the order of $10^{-2}$. 
In summary, a compactness of order ${\cal C} \gtrsim 10^{-2}$ can be achieved for $0 < N_0 \lesssim 0.7$ by adopting the criterion
$0.90M \le {\cal M}(r_s) 
\le 0.99M$ for determining the radius of the object.

\section{Purely magnetic objects} 
\label{magneticsec}

We now proceed to study the purely magnetic object, characterized by
\be
q_M \neq 0\,,\qquad 
q_E=0\,,
\ee
under which $A_0'(r) = 0$, 
as follows from Eq.~(\ref{back3}).
Using Eq.~(\ref{back2}) with the branch $\phi'(r) > 0$, we obtain $\phi'(r)$ in the same form as in Eq.~(\ref{rphi}).
The corresponding energy density of the scalar field, $\rho_{\phi} 
=h \phi'^2/2$, is given by 
\be
\rho_{\phi}=
\frac{\Mpl^2 hN'}{2rN}\,,
\label{rhophiM}
\ee
which has the same form as 
Eq.~(\ref{rhophi}) derived for the electric case.
Substituting the energy 
density $\rho=h \phi'^2/2+\mu q_M^2/(2r^4)$ into the background 
Eq.~(\ref{back1}), the 
coupling $\mu$ can be expressed as
\be
\mu=-\frac{\Mpl^2 r^2 [2(rh'+h-1)N+rhN']}
{q_M^2 N}\,.
\label{mum}
\ee
We introduce the dimensionless magnetic charge
\be
\tilde{q}_M \equiv 
\frac{q_M}{\Mpl r_0}\,,
\ee
and define 
\be
\mu_M \equiv \tilde{q}_M^2 
\mu = {\cal F}^{-1}\,,
\ee
where ${\cal F}$ is the function given in 
Eq.~(\ref{calF}).
Compared with the coupling (\ref{muE}) in the electric case, the following relation holds:
\be
\mu_E \mu_M = 1\,,
\label{muEM}
\ee
which demonstrates the 
electric-magnetic duality.
The energy density associated with the magnetic charge, 
$\rho_M=\mu q_M^2/(2r^4)$, 
is given by 
\be
\rho_{M}=-\frac{\Mpl^2
[2(rh'+h-1)N+rhN']}{2r^2 N}\,,
\label{rhoM}
\ee
which is of the same form as Eq.~(\ref{rhoA02}) in 
the electric case.

From Eq.~(\ref{back4}), the 
$\phi$ derivative of $\mu$ 
can be expressed 
in the form:
\be
\mu_{,\phi}=\frac{2r^2}
{q_M^2 \sqrt{N}} \left( 
\sqrt{N} h r^2 \phi' 
\right)'\,.
\label{mupm}
\ee
We differentiate Eq.~\eqref{mum} with respect to $r$ and use the relation $\mu' = \mu_{,\phi}\,\phi'$.
Substituting Eq.~\eqref{mupm} together with $\phi' = \Mpl \sqrt{N'/(rN)}$ 
into the right-hand side of this relation, we obtain
\ba
& &
2 \left( r^2 h''+4r h'+2h-2 
\right)N^2-r^2 h N'^2 
\nonumber \\
& &
+3r( rh'+2h) NN'
+2r^2 h N N''=0\,.
\label{hNeq2}
\ea
which is the same form as Eq.~(\ref{hNeq}) derived for the electric case. 
For a given function $N(r)$, the integrated solution for $h(r)$ can be found in the form of Eq.~(\ref{hr2}). 
From Eqs.~\eqref{rhophiM} and \eqref{rhoM}, together with Eqs.~\eqref{rhoD}-\eqref{PtD}, 
the energy density $\rho = \rho_{\phi} + \rho_{M}$ 
and the pressures $P_r$ and $P_t$ can be expressed in terms of the metric functions 
$N(r)$ and $h(r)$, as well as their derivatives.
Thus, for a given $N(r)$, we obtain the same solutions to $h(r)$, $f(r)$, ${\cal M}(r)$, 
$\phi'(r)$, 
$\rho(r)$, $P_r(r)$, and 
$P_t (r)$ 
as those in the 
electric case. 

For the choice of $N(r)$ given in Eq.~(\ref{Nexample}), 
the metric function $h(r)$ is analytically obtained in the form of Eq.~(\ref{hana}), 
together with the mass
function ${\cal M}(r)$ given 
in Eq.~(\ref{mfunction}).
The solution to the scalar field is given by Eq.~(\ref{phiso}), where $\phi_0$ 
is the field value at $r=0$.
The field value $\phi_{\infty}$ at spatial 
infinity is related to $\phi_0$ according to the 
relation (\ref{phiinf}). 
For $N_0 = 0.5$, the behaviors of 
$f$, $h$, ${\cal M}$, 
$\phi'$, $\phi - \phi_0$, 
$\rho$, $P_r$, $\hat{\rho}$, and  
$\hat{P}_r$ are the same as those 
shown in Figs.~\ref{figmet}, \ref{figAphi}, and \ref{figrhoP}, but without the electric field ($A_0'(r) = 0$). 

\vspace{0.5cm}
\begin{figure}[ht]
\begin{center}
\includegraphics[height=3.2in,width=3.4in]{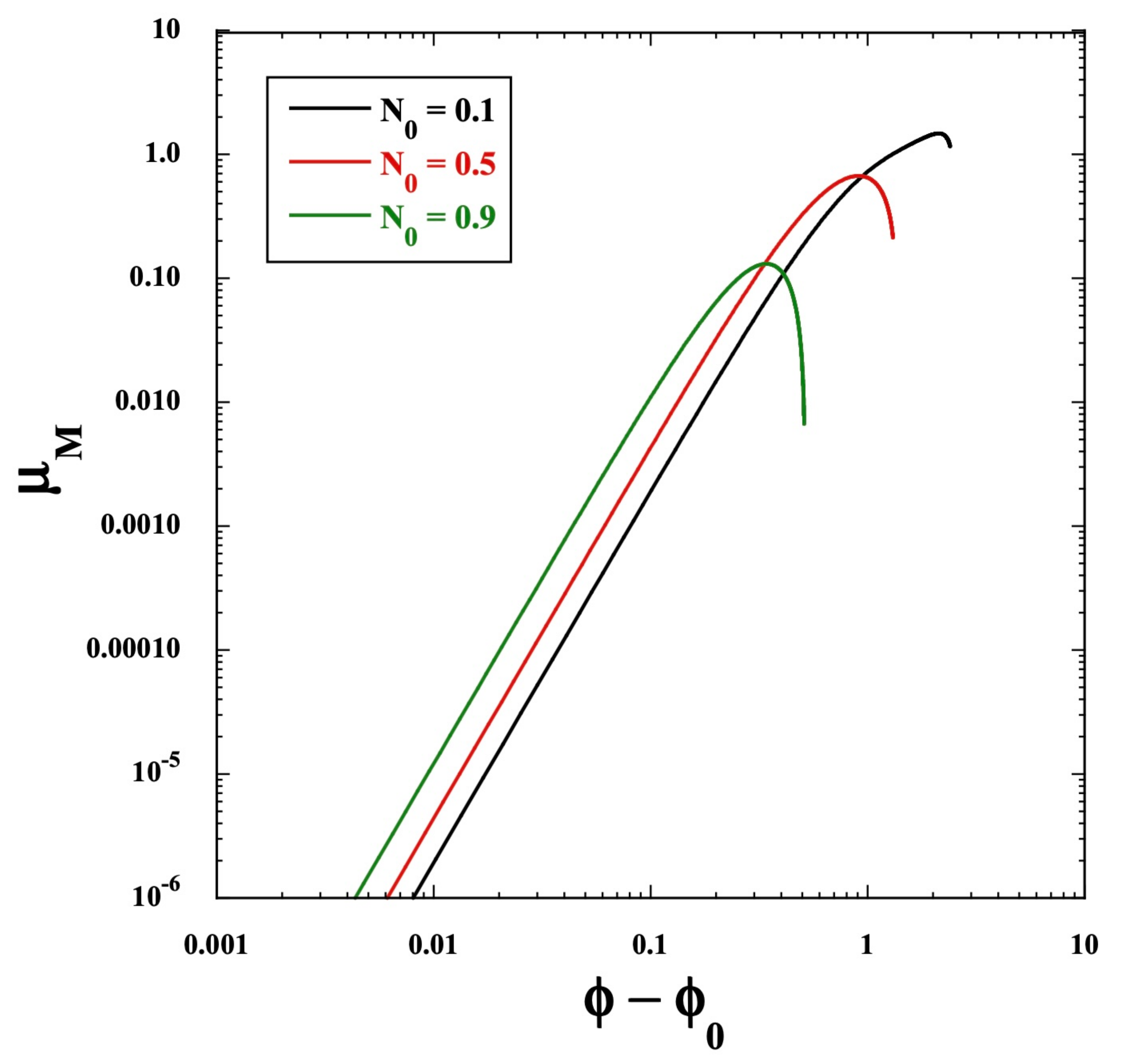}
\end{center}
\caption{The rescaled coupling 
$\mu_M=\tilde{q}_M^{2}\mu$ as a function of 
$\phi-\phi_0$ (normalized by $\Mpl$) 
with three different values of $N_0$.
\label{figmuM}}
\end{figure}

As we showed in Eq.~(\ref{muEM}), the coupling $\mu_M$ is the inverse of 
$\mu_E$ defined in Eq.~(\ref{muE}). 
Around the center of the body, the coupling 
has the following dependence:
\be
\mu_M (\phi) \simeq 
\frac{4(\phi-\phi_0)^3}
{\Mpl^3 \lambda}\,,
\label{mum0}
\ee
at leading order. Since the scalar field 
behaves as Eq.~(\ref{phiso2}) near $r=0$, 
Eq.~(\ref{mum0}) translates to the 
radial dependence $\mu_M=4 \lambda^2 x^6 
\propto r^6$. 
At large distances, the coupling has 
the following $\phi$ dependence:
\be
\mu_M (\phi) \simeq 
\frac{\pi^2 \lambda^4}
{4(\lambda^2+2)^2}
\left[ 1-\frac{16 \sqrt{2}
(\lambda^2+2)^{3/2}
(\phi-\phi_{\infty})}
{\pi^2 \Mpl \lambda^3} \right]\,,
\label{muMinf}
\ee
where  $\phi-\phi_{\infty} 
\simeq -\sqrt{2}\Mpl 
\lambda/(\sqrt{\lambda^2+2}
\,x^2)$. Thus, $\mu_M$ decreases as 
a function of $\phi$ (or $r$) toward 
its asymptotic value 
\be
\mu_{M,\infty}=\frac{\pi^2 \lambda^4}{4(\lambda^2+2)^2}
=\frac{\pi^2 (1-\sqrt{N_0})^2}{4}\,.
\label{muMinf2}
\ee

In Fig.~\ref{figmuM}, we plot $\mu_M$ as a function of $\phi-\phi_0$ for $N_0=0.1, 0.5, 0.9$.
The coupling $\mu_M(\phi)$ increases proportionally to $(\phi-\phi_0)^3$ in the regime $\phi-\phi_0 \ll \Mpl$, and then reaches a maximum.
It subsequently decreases toward its asymptotic value $\mu_{M,\infty}$. 
As estimated from Eq.~(\ref{muMinf2}), $\mu_{M,\infty}$ decreases with increasing $N_0$. 
Comparing Fig.~\ref{figmuM} with Fig.~\ref{figmuE}, 
the theoretical curve of $\mu_M(\phi)$ corresponds to the reciprocal of $\mu_E(\phi)$.

With the choice (\ref{Nexample}) for $N(r)$, the energy density profile $\rho(r)$ for a given value of $N_0$ is identical to that shown in Fig.~\ref{figrhoN0}.
The relation between the ADM mass $M$ and the radius $r_s$ is also the same as that of the electric configuration plotted in Fig.~\ref{figMr}. 
The ADM mass is analytically given by Eq.~(\ref{ADM2}) as a function of $N_0$. Defining the radius $r_s$ by the condition ${\cal M}(r_s)=0.99M$, we obtain the mass-radius relation shown as the solid line in Fig.~\ref{figMr}.
Relaxing the condition to ${\cal M}(r_s)=0.90M$ shifts the $(M, r_s)$ relation to the dashed line, yielding smaller radii $r_s$ and larger compactness ${\cal C}$.

\section{Perturbations and 
strong coupling issue} 
\label{obsersec}

In theories described by the action (\ref{action}),
the second-order action of perturbations around the background (\ref{metric})
was already derived in Ref.~\cite{DeFelice:2024ops}.
In that work, the authors also obtained the conditions for the absence of ghosts and Laplacian instabilities by taking the 
small-scale limit.
In ESM theories where the matter sector is characterized by the Lagrangian
${\cal L}= X + \mu(\phi)F$,
the theory is ghost-free for $\mu(\phi) > 0$.
Indeed, as shown in Figs.~\ref{figmuE} and \ref{figmuM},
the coupling $\mu$ remains positive for both electric and magnetic configurations.
Moreover, the propagation speeds of the five dynamical degrees of freedom
coincide with the speed of light \cite{DeFelice:2024ops}, ensuring the absence of Laplacian instabilities.

Thus, both the electric and magnetic compact objects realized in ESM theories
are free from ghost and Laplacian instabilities.
However, there remains one issue that was not 
addressed in Ref.~\cite{DeFelice:2024ops}.
For the compact objects 
studied in Secs.~\ref{electricsec} 
and \ref{magneticsec}, 
the coupling $\mu (\phi)$ changes 
rapidly near $r=0$. 
In particular, when $\mu$ approaches 0 toward $r=0$,
as in the case of the magnetic object,
this may indicate the emergence of a strong coupling problem
(see Ref.~\cite{Demozzi:2009fu} for a similar situation arising in inflationary magnetogenesis).
In the following, we address the issue of strong coupling.

We write the perturbations of 
the metric tensor 
$g_{\mu \nu}$ on the background 
(\ref{metric}), as $h_{\mu \nu}$, and then expand each 
component of them in terms of the spherical harmonics 
$Y_{lm}(\theta, \varphi)$ 
\cite{Regge:1957td, Zerilli:1970se, Moncrief:1974ng, Zerilli:1974ai}. 
We consider the mode $m=0$ 
without loss of generality 
and write $Y_{l0}$ as $Y_{l}$ 
for simplicity.
By choosing the gauge conditions 
$h_{t \theta}=0$, $h_{\theta \theta}=0$, 
$h_{\varphi \varphi}=0$, and 
$h_{\theta \varphi}=0$, the 
nonvanishing components of $h_{\mu \nu}$ are given by \cite{DeFelice:2011ka,Kobayashi:2012kh,Kobayashi:2014wsa,Kase:2021mix}
\ba
\hspace{-0.4cm}
& &
h_{tt}=f(r) H_0 (t,r) Y_{l}(\theta), \quad 
h_{tr}=H_1 (t,r) Y_{l}(\theta),
\nonumber \\
\hspace{-0.4cm}
& &
h_{t \varphi}=-Q(t,r) (\sin \theta) 
Y_{l, \theta} (\theta),
\;\; 
h_{rr}=h^{-1}(r) H_2(t,r) Y_{l}(\theta),\nonumber \\
\hspace{-0.4cm}
& &
h_{r \theta}=
h_1 (t,r)Y_{l, \theta}(\theta),
\quad
h_{r \varphi}=-W(t,r) (\sin \theta) Y_{l,\theta} (\theta), 
\nonumber \\
\hspace{-0.4cm}
\label{hcom}
\ea
where $H_0$, $H_1$, $H_2$, $h_1$, $Q$, and $W$ depend on $t$ and $r$. 
We also decompose the scalar and vector fields, as 
\ba
\phi &=& 
\bar{\phi}(r)+\delta \phi (t,r) Y_{l}(\theta),\\
A_{\mu} &=& 
\bar{A}_{\mu}(r)
+\delta A_\mu\,,
\label{perma}
\ea
where
\ba
& &
\delta A_t=\delta A_0 (t,r) Y_{l}(\theta),\qquad 
\delta A_r=\delta A_1 (t,r) Y_{l}(\theta),\qquad \nonumber \\
& &
\delta A_\theta=0,\qquad 
\delta A_{\varphi}=
-r \widetilde{\delta A}(t,r) (\sin \theta) 
Y_{l,\theta}(\theta)\,.
\ea
The existence of the $U(1)$ gauge invariance allows us to set $\delta A_\theta = 0$. 
Moreover, we multiplied 
$\widetilde{\delta A}(t,r)$ by $r$ so that 
$\delta A_{\varphi}$ can be written as  $A_\mu{\rm d}x^\mu\ni -\widetilde{\delta A}(t,r) Y_{l,\theta}(\theta) (r\sin\theta\,{\rm d}\varphi)$, where $r\sin\theta\,{\rm d}\varphi$ is the orthonormal basis covector. 
Compared with the notation $\delta A$ used in Ref.~\cite{DeFelice:2024ops}, the variables are related by 
$\delta A = r \widetilde{\delta A}$.
The normalization with respect to 
$\widetilde{\delta A}$ is necessary 
when studying whether the solutions are strongly coupled.

In the odd-parity sector, the 
gauge-invariant combination of 
the form $\chi=r\dot{W}-rQ'+2Q 
-2{\cal L}_{,F}A_0' r \delta A/\Mpl^2$
was introduced in Ref.~\cite{DeFelice:2024ops}, where 
a dot represents the derivative 
with respect to $t$.
Analogous to $\delta A_{\varphi}$, 
we can express $h_{t \varphi}$ as
$h_{t \varphi}=-(Q/r)(r\sin \theta) 
Y_{l, \theta} (\theta)$.
The normalized field $\tilde{\chi}$ suitable for studying the issue of strong coupling should contain the term $Q/r$.
Therefore, we introduce 
$\tilde{\chi}=\chi/r$, 
that is
\be
\tilde{\chi} =
\dot{W}-Q'+\frac{2Q}{r}
-\frac{2{\cal L}_{,F} r A_0' \widetilde{\delta A}}{\Mpl^2}\,.
\ee
We also have the normalized 
odd-parity perturbation 
$\widetilde{\delta A}=\delta A/r$ arising from the covector field.

In the even-parity 
sector, Ref.~\cite{DeFelice:2024ops} introduced a gauge-invariant 
combination arising from 
the gravitational sector, 
given by $\psi = 
r H_2 - L h_1$. 
Since no additional factor of $r$ is required for the proper normalization 
of $H_2$, we define the rescaled field $\tilde{\psi} = \psi/r$, namely,
\be
\tilde{\psi}=H_2-\frac{L h_1}{r} \,.
\ee
For the vector-field perturbation, 
we introduce the gauge-invariant combination  
\be
\tilde{V}=r \left[ \delta A_0'-\dot{\delta A}_1
+\frac{A_0'}{2} \left( 
H_0-H_2 \right)
+\frac{A_0' \mu_{,\phi}}{\mu} \delta \phi 
\right]\,,
\ee
which is related to the variable $V$ introduced 
in Ref.~\cite{DeFelice:2024ops} through the relation $\tilde{V}=rV$.
Defining $\tilde{V}$ in this way 
ensures that the vector-field perturbation remains regular 
at spatial infinity.
There is also a scalar-field perturbation 
$\delta \phi$ as a dynamical 
propagating degree of freedom 
in the even-parity sector.

\subsection{Electric objects}

Let us first consider the case of electric objects. 
Expanding the action (\ref{action}) up to quadratic 
order in perturbations with the Lagrangian (\ref{model}), 
the reduced second-order action takes the form 
\ba
\hspace{-0.7cm}
{\cal S}^{(2)} &=& 
\int {\rm d}t{\rm d}r\, r^2\sqrt{\frac{f}{h}}\, 
[ K_{ab} \dot{\cal X}_a
\dot{\cal X}_b
-M_{ab} {\cal X}_a  
{\cal X}_b \nonumber \\
\hspace{-0.7cm}
& & +\tilde{K}_{ij} \dot{\cal Y}_i 
\dot{\cal Y}_j
-\tilde{M}_{ij}{\cal Y}_i 
{\cal Y}_j
-\tilde{B}_{ij} (\dot{\cal Y}_i
{\cal Y}_j-{\cal Y}_i\dot{\cal Y}_j)]\,,
\label{S2}
\ea
where the indices $a,b \in {1,2}$ label the odd-parity 
modes ${\cal X}_a \in \{ \tilde{\chi}, \widetilde{\delta A} \}$, 
and the indices 
$i,j \in \{ 1,2,3 \}$ label the even-parity modes 
${\cal Y}_i \in \{ \tilde{\psi}, \delta\phi, \tilde{V} \}$. 
The matrices $K$, $M$, $\tilde K$, and $\tilde M$ are real 
and symmetric, whereas the matrix $\tilde{B}$ is real but antisymmetric. 
In the square bracket 
of Eq.~(\ref{S2}), 
the summation is taken over all the components $a,b$ and $i,j$. 
We also note that the background determinant, excluding the angular part which has already been integrated,\footnote{Using the coordinate $z=\cos\theta$, we have exactly $\sqrt{-g} = r^2 \sqrt{f/h}$.}
$\sqrt{-g} = r^2 \sqrt{f/h}$, appears explicitly in Eq.~(\ref{S2}).
This prescription is important for correctly addressing 
the issue of strong coupling.

The matrix $K$, which is associated with the odd-parity perturbations, has only diagonal components.  
The corresponding kinetic part of the Lagrangian 
${\cal L}_K= r^2\sqrt{f/h}\, (K_{11}\dot{\tilde\chi}^2+K_{22}\dot{\widetilde{\delta A}}{}^2)$ can be written as 
${\cal L}_K=r^2\sqrt{f/h} [K_{11}f\,(\dot{\tilde\chi}/\sqrt{f})^2+K_{22}f\,(\dot{\widetilde{\delta A}}/\sqrt{f})^2]$.
In this form, we are considering the 
partial derivative $(1/\sqrt{f})\partial/\partial t$, where $\sqrt{f} {\rm d}t$ represents the orthonormal infinitesimal timelike interval, since $\sqrt{f}$ is the lapse function.\footnote{This is analogous to the normalization usually performed 
for the kinetic Lagrangian  $\mathcal{L}_K=\sqrt{-g}\, Q(\dot{\psi}/\mathcal{N})^2$ on an isotropic cosmological background. 
Here, $\psi$ denotes a perturbed field, and 
$\sqrt{-g} = \mathcal{N} a^3$, 
where $a$ is the scale factor and $\mathcal{N}$ is the lapse function. 
The time-dependent function $Q$ represents the normalized kinetic coefficient, 
for which the strong coupling regime 
corresponds to the limit $Q \to 0$.} 
Therefore, after multiplying 
$K_{11}$ and $K_{22}$ by $f$, 
we obtain
\begin{align}
K_{11}f&=\frac{L\Mpl^2}{4N (L-2)}\,,\\ 
K_{22}f&=\frac{L \mu}{2}\,,
\end{align}
where $L=l(l+1)$. 
The positivity of $K_{11}f$ 
characterizes the ghost-free condition for the gravitational perturbation $\tilde{\chi}$. 
Since $K_{11}f$ is always positive, there is no ghost.
At $r = 0$, we have $K_{11}f = L \Mpl^2/[4N_0 (L - 2)]$, 
while at spatial infinity 
$K_{11}f \to L \Mpl^2/[4(L - 2)]$.
Since $K_{11}f$ does not approach 0 
at any radius $r$, 
the perturbation $\tilde{\chi}$ is not strongly coupled.

The other component, $K_{22} f$, remains positive for $\mu > 0$, ensuring that the vector-field perturbation $\widetilde{\delta A}$ is ghost-free.
Near $r = 0$, we have $\mu \propto r^{-6}$, and hence $K_{22}f$ also increases proportionally to $r^{-6}$.
This behavior corresponds to a weak-coupling limit for the vector-field perturbation $\widetilde{\delta A}$. 
This is expected, since for this solution 
$\mu(\phi) \to +\infty$, and $\mu(\phi)$ multiplies the kinetic term $F$.
At spatial infinity, $K_{22}f$ 
approaches a positive constant. 
The odd-parity vector-field perturbation is weakly coupled near the center, with no regime where $\widetilde{\delta A}$ becomes strongly coupled.

To investigate the strong coupling issue for even-parity perturbations, 
we need to diagonalize the kinetic matrix $\tilde{K}_{ij}$.
One of the simplest ways to achieve this is by performing a field 
redefinition, 
${\cal Y}_i = T_{ij} \bar{\cal Y}_j$, 
where $T$ is a lower unitriangular matrix. 
In particular, this transformation leads to 
$\tilde{\psi}={\cal Y}_1=\bar{\cal Y}_1$, 
and $\bar{K}_{33}=(T^t \tilde{K}T)_{33}=\tilde{K}_{33}$. 
Such a transformation always exists 
because $\det(T) = 1$. 
Upon performing this step and taking 
the $l \gg 1$ limit, we find that 
the three diagonal elements are given by 
\ba
\bar{K}_{11}f&=&\frac{\det(\tilde{K})f}
{\tilde{K}_{22} \tilde{K}_{33}-\tilde{K}_{23}^2}
= \frac{\Mpl^2 h^2}{L^2}\,,\\
\bar{K}_{22}f&=&\frac{(\tilde{K}_{22} 
\tilde{K}_{33}-\tilde{K}_{23}^2)f}{\tilde{K}_{33}}
=\frac{1}{2}\,,\\
\bar{K}_{33}f &=&\tilde{K}_{33}f
=\frac{\mu}{2LN}\,,
\ea
which are all positive for $\mu>0$. 
As in the case of $K_{22}f$ for 
odd-parity modes, we have 
$\bar{K}_{33}f \to \infty$ for $r \to 0$ 
due to the behavior $\mu \propto r^{-6}$.
This shows that the vector-field perturbation $\tilde{V}$ in the even-parity sector is weakly coupled near $r = 0$. 
At spatial infinity, 
$\bar{K}_{33}f$ approaches 
a nonvanishing constant. 
The strong coupling problem does not arise for $\tilde{V}$ at any radius $r$. 
The other components, 
$\bar{K}_{11}f$ and $\bar{K}_{22}f$, characterize the kinetic behavior of the remaining two dynamical perturbations in the even-parity sector.
Since both $\bar{K}_{11}f$ and $\bar{K}_{22}f$ approach constant values in the limits $r \to 0$ and $r \to \infty$, there are no strong coupling issues for $\tilde{\psi}$ and $\delta \phi$. 

Since $\bar{K}_{22} f = 1/2$, the field $\bar{\mathcal{Y}}_2$ is already canonically normalized. 
For the remaining two fields, 
the positivity of the normalized 
kinetic terms allows one to perform the final field redefinitions.
By introducing $\bar{\mathcal{Y}}_1=
2^{-1/2}L \hat{\mathcal{Y}}_1/(\Mpl h)$ and $\bar{\mathcal{Y}}_3
=\sqrt{L N} \hat{\mathcal{Y}}_3/\sqrt{\mu}$, 
we obtain the canonical kinetic terms 
for $\hat{{\cal Y}}_1$ and 
$\hat{{\cal Y}}_3$, respectively.
The fact that $\mu \to +\infty$ 
at the origin implies that 
terms higher than second order in 
perturbations are infinitely suppressed, 
leading to a decoupling behavior 
of the vector-field perturbation $\tilde{V}$. 
We have thus shown that electric objects are free from strong coupling problems in both the 
odd- and even-parity sectors.

\subsection{Magnetic objects}

We now proceed to the case of magnetic 
objects. After removing the nondynamical 
perturbations, the reduced second-order action can be expressed in the form 
\begin{align}
{\cal S}^{(2)} &= 
\int {\rm d}t{\rm d}r\, r^2\sqrt{\frac{f}{h}}\, 
[ {\cal K}_{ab} \dot{\cal Z}_a
\dot{\cal Z}_b-{\cal B}_{ab} (\dot{\cal Z}_a
{\cal Z}_b-{\cal Z}_a\dot{\cal Z}_b)
 \nonumber \\
& \quad
-{\cal M}_{ab} {\cal Z}_a  
{\cal Z}_b+\tilde{\cal K}_{ij} \dot{\cal W}_i 
\dot{\cal W}_j
-\tilde{\cal M}_{ij}{\cal W}_i 
{\cal W}_j\nonumber\\
& \quad 
-\tilde{\cal B}_{ij} (\dot{\cal W}_i
{\cal W}_j-{\cal W}_i\dot{\cal W}_j)]\,,
\label{S2_mag}
\end{align}
where ${\cal Z}_a\in\{\tilde{\chi}, \tilde{V} \}$ and ${\cal W}_i\in\{\tilde\psi,\delta\phi,
\widetilde{\delta A} \}$. All matrices are real and symmetric, except for ${\cal B}$ and $\tilde{\cal B}$, which are antisymmetric. 
Note that ${\cal Z}_a$ consists of 
the odd-parity gravitational perturbation $\tilde{\chi}$ 
and the even-parity vector-field perturbation $\tilde{V}$, 
while ${\cal W}_i$ contains the 
odd-parity vector-field perturbation $\widetilde{\delta A}$ 
in addition to the even-parity modes $\tilde{\psi}$ 
and $\delta\phi$.

The $2 \times 2$ matrix ${\cal K}$, 
associated with the kinetic terms of the fields ${\cal Z}_a$, has both diagonal and off-diagonal components.
We can diagonalize ${\cal K}$ by using 
a lower unitriangular matrix. 
Taking the $l \gg 1$ limit, we find 
that two diagonal components are 
given by 
\ba
& &
\bar{\cal K}_{11}f =
\frac{({\cal K}_{11}
{\cal K}_{22}-{\cal K}_{12}^2)f}{{\cal K}_{22}}
=\frac{\Mpl^2}{4N}\,,
\label{K11}\\
& &
\bar{\cal K}_{22}f=
{\cal K}_{22}f=\frac{\mu}{2LN}\,.
\label{K22}
\ea
We also perform a similar procedure 
for the $3 \times 3$ matrix 
$\tilde{{\cal K}}$, which is
associated with the fields 
${\cal W}_i$.
In the limit $l \gg 1$, we obtain 
the following three diagonal components:
\ba
\overline{\tilde{\cal K}}_{11} f
&=&
\frac{\det(\tilde{\cal K})f}{\tilde{\cal K}_{22}\tilde{\cal K}_{33}-\tilde{\cal K}_{23}^2}=\frac{\Mpl^2 h^2}{L^2}\,,\\
\overline{\tilde{\cal K}}_{22} f
&=& \frac{(\tilde{\cal K}_{22}\tilde{\cal K}_{33}-\tilde{\cal K}_{23}^2)f}{\tilde{\cal K}_{33}}
=\frac{1}{2}\,,\\
\overline{\tilde{\cal K}}_{33} f
&=& 
\tilde{\cal K}_{33}f=\frac{L\mu}{2}\,.
\label{cK33}
\ea
All of Eqs.~(\ref{K11})-(\ref{cK33}) are positive for $\mu>0$, so that no ghosts arise in these solutions. 
However, both Eqs.~(\ref{K22}) and (\ref{cK33}) are proportional to $\mu$.
For magnetic objects studied in Sec.~\ref{magneticsec}, 
the coupling $\mu$ behaves as $\mu \propto r^6$ near the center. 
This leads to vanishing kinetic coefficients for $\tilde{V}$ and $\widetilde{\delta A}$ at $r=0$, 
so that these solutions are strongly coupled around the origin. 

In other words, upon canonical normalization of the kinetic terms for $\tilde{V}$ and $\widetilde{\delta A}$, 
terms higher than second order in the perturbative expansion of the Lagrangian acquire $\sqrt{\mu}$-dependent contributions 
in their denominators. 
As a result, the nonlinear terms dominate over the linear contributions 
in the limit $\mu \to 0$. 
In this regime, the analysis based on linear perturbations loses its validity near $r=0$, 
signaling the breakdown of the effective field theory.

In summary, the magnetic compact object, which is realized through the coupling $\mu$ in Eq.~(\ref{mum}), is prone to the 
strong coupling problem around the center. 
This problem originates from the divergent behavior of $F$, which behaves as 
$F = -q_M^2/(2r^4)$ at the center.  
In this case, the coupling $\mu$ must offset the growth of $F$ so that the product $\mu F$ remains finite at $r = 0$.
Since the coupling depends as 
$\mu(\phi) \propto (\phi - \phi_0)^3 \propto r^6$, 
the system inevitably enters a strong coupling regime. 

For electric objects, however, the field strength $F$ behaves near the center as $F \propto r^8$, so that the product $\mu F$ remains finite even if $\mu$ diverges at $r=0$.
In this case, the coupling behaves as $\mu(\phi) \propto (\phi - \phi_0)^{-3} \propto r^{-6}$, corresponding to a weakly coupled regime. 
Although the same energy density, pressure, mass, 
and radius can be realized for both electric and magnetic objects, 
the two models are physically distinct due to the different 
functional dependences of the coupling $\mu(\phi)$. 
We have shown that only the electric object remains free from the strong coupling problem.

\section{Conclusions} 
\label{consec}

We studied the existence of regular ECOs without central singularities in a class of scalar-vector-tensor theories. 
In k-essence theories within 
the framework of GR, the absence of ghosts forbids static SSS solutions with a positive-definite energy density. By introducing the dependence of the gauge-field strength $F$ in the k-essence Lagrangian, we showed that regular ECOs with positive-definite energy can be achieved without ghosts.

In particular, we focused on ESM theories described by the Lagrangian $\Mpl^2/2+X+\mu(\phi)F$. 
We introduced the background density $\rho$ and the radial and angular pressures $P_r$ and $P_t$, as defined in Eqs.~(\ref{rho})-(\ref{Pt}), to investigate the density profiles and the balance between the pressure gradient and gravity. 
As seen from the continuity equation (\ref{continuity2}), as long as ${\cal L}_{,F} = \mu(\phi) > 0$, the pressure gradient induced by the coupling between the scalar and vector fields can counteract gravity.

For the electric object, we showed that the metric function $h(r)$ can be written in the integrated form (\ref{hr2}), where $N(r) = f(r)/h(r)$. 
Since $h(r) > 0$ everywhere, the solution does not correspond to a BH but to an ECO without horizons.
For the choice (\ref{Nexample}), we were able to obtain an analytic solution for $h(r)$, given in Eq.~(\ref{hana}), so that the background metric components are explicitly known. 
Accordingly, the scalar-field derivative $\phi'$, the electric field $A_0'$, the coupling $\mu$, and the quantities $\rho$, $P_r$ $(=-P_t)$ are all determined analytically.
We showed that $\rho$ vanishes at the center, with the nonrelativistic equation of state $w_r = P_r/\rho \to 0$. 
The energy density reaches a maximum around the radius $r = r_0$ and then decreases toward 0 at spatial infinity.
The peculiar feature of the vanishing energy density and pressures is attributed to the fact that the form of $N(r)$ is constrained by the boundary condition $\phi'(0) = 0$.

The coupling $\mu$ for the electric ECO is given by Eq.~(\ref{mu2}). Near $r = 0$, $\mu$ behaves as $\mu \propto (\phi - \phi_0)^{-3}$, while at large distances its behavior is described by Eq.~(\ref{muasy}), so that $\mu(\phi)$ approaches a constant value $\mu_{\infty}$. 
The coupling exhibits a minimum at an intermediate distance, with a positive value for any $N_0$ in the range $0 < N_0 < 1$. Therefore, the ghost-free condition $\mu(\phi) > 0$ is satisfied at all distances $r$. 

We also found that the ADM mass of the electric object is given by $M=2\sqrt{2}\pi^2 (1-\sqrt{N_0})M_0$, where $M_0=\Mpl^2 r_0$. For a given $r_0$, $M$ has a maximum value $M_{\rm max}=2\sqrt{2}\pi^2 \Mpl^2 r_0$ in the limit $N_0 \to 0$. 
We defined the radius $r_s$ of the object as the point where the mass function reaches 99\,\% of the ADM mass, i.e., 
${\cal M}(r_s) = 0.99 M$. 
In this case, the maximum radius reached in the limit $N_0 \to 0$ is $(r_s)_{\rm max}=55.6r_0$. Since $r_0$ can take any positive value, both 
$M_{\rm max}$ and $(r_s)_{\rm max}$ can be arbitrarily large, while the ratio 
$M_{\rm max}/(r_s)_{\rm max}$ 
is fixed.
In Fig.~\ref{figMr}, we plot the mass-radius relation as a solid line, showing that both $M$ and $r_s$ decrease with increasing $N_0$. The compactness of the object is approximately 0.02 for $0 < N_0 \lesssim 0.7$. 
Using an alternative criterion, ${\cal M}(r_s) = 0.90 M$, to define the radius, the compactness can reach the order of 0.1.

For the magnetic ECO, the metric functions satisfy the differential equation (\ref{hNeq2}), whose integrated solution coincides with Eq.~(\ref{hana}) in the electric case. 
Consequently, with the choice of $N(r)$ given by Eq.~(\ref{Nexample}), the resulting forms of $h(r)$ and $f(r)$ are the same as those obtained in the electric configuration. 
Indeed, the functional dependence of $\phi'$, as well as that of $\rho$ and $P_r$ $(=-P_t)$, is identical in both the electric and magnetic cases. 
The difference appears only for the coupling $\mu(\phi)$, where the rescaled magnetic coupling $\mu_M=\tilde{q}_M^2 \mu$ is related to the rescaled 
electric coupling $\mu_E=\tilde{q}_E^{-2} \mu$, as $\mu_M=\mu_E^{-1}$. Near the center of the body, the 
coupling behaves as $\mu_M \propto (\phi - \phi_0)^3$, approaching 0 as $r \to 0$.
At spatial infinity, $\mu_M$ approaches the constant value $\pi^2 (1 - \sqrt{N_0})^2/4$. 
As shown in Fig.~\ref{figmuM}, $\mu_M(\phi)$ reaches a maximum at an intermediate field value. 

For both the electric and magnetic objects, neither ghost nor Laplacian instabilities arise at any radius $r$. 
In this paper, we addressed the issue of strong coupling by properly introducing normalized dynamical perturbations on the SSS background. 
We showed that the electric object is free from this problem and that the vector-field perturbations remain weakly coupled near the center.
On the other hand, the magnetic object suffers from the strong coupling problem of vector-field perturbations, as $\mu(\phi)$ approaches 0 in the limit 
$r \to 0$. 
Thus, only the electric configuration can serve as a viable candidate for an ECO without theoretical pathologies.

If a binary system containing at least one electric ECO exists, it should be possible to probe its signatures through gravitational wave observations. 
During the inspiral phase of a binary coalescence, the gravitational waveform is expected to be influenced by the presence of electric charges associated with dark photons.
As in the case of a scalar charge \cite{Alsing:2011er,Yunes:2011aa,Berti:2018cxi,Higashino:2022izi,Takeda:2023wqn}, the observed phase of the gravitational waveform should allow one to place bounds on the magnitude of the dark electric charge. 
Moreover, the Love number of ECOs is expected to differ from that of NSs \cite{Flanagan:2007ix,Hinderer:2007mb} due to the peculiar behavior of the density profile near the center. 
It could be constrained by future observations of the tidal deformability in binary systems containing ECOs. It would also be interesting to investigate whether such a compact-object configuration can arise from the collapse of scalar- and vector-field energy densities. 
A detailed analysis of these intriguing issues is left for future work.

\section*{Acknowledgements}

We are grateful to Nathalie Deruelle 
and Carlos Herdeiro for useful correspondence. 
ST is supported by JSPS KAKENHI Grant Number 22K03642 and Waseda University Special Research 
Projects (Nos.~2025C-488 and 2025R-028).

\bibliographystyle{mybibstyle}
\bibliography{bib}

\end{document}